\documentclass[reprint,superscriptaddress,amsmath,amssymb,aps,longbibliography,floatfix]{revtex4-1}

\usepackage{mathtools}
\usepackage[colorlinks]{hyperref}
\usepackage{bm}
\usepackage{algorithm}
\usepackage{algpseudocode}

\usepackage{tikz}

\newcommand{\eq}[1]{Eq.~\hyperref[eq:#1]{(\ref*{eq:#1})}}
\renewcommand{\sec}[1]{\hyperref[sec:#1]{Section~\ref*{sec:#1}}}
\newcommand{\app}[1]{\hyperref[app:#1]{Appendix~\ref*{app:#1}}}
\newcommand{\tab}[1]{\hyperref[tab:#1]{Table~\ref*{tab:#1}}}
\newcommand{\fig}[1]{\hyperref[fig:#1]{Figure~\ref*{fig:#1}}}
\newcommand{\figa}[2]{\hyperref[fig:#1]{Figure~\ref*{fig:#1}#2}}
\newcommand{\figx}[2]{\hyperref[fig:#1]{Figure~\ref*{fig:#1}(#2)}}
\newcommand{\thm}[1]{\hyperref[thm:#1]{Theorem~\ref*{thm:#1}}}
\newcommand{\lem}[1]{\hyperref[lem:#1]{Lemma~\ref*{lem:#1}}}
\newcommand{\cor}[1]{\hyperref[cor:#1]{Corollary~\ref*{cor:#1}}}
\newcommand{\defn}[1]{\hyperref[def:#1]{Definition~\ref*{def:#1}}}
\newcommand{\alg}[1]{\hyperref[alg:#1]{Algorithm~\ref*{alg:#1}}}

\newcommand{\be}{\begin{equation}}
\newcommand{\ee}{\end{equation}}
\newcommand{\ba}{\begin{eqnarray}}
\newcommand{\ea}{\end{eqnarray}}

\DeclarePairedDelimiter\set{\{}{\}}
\DeclarePairedDelimiter\parens{(}{)}

\DeclarePairedDelimiter\bracks{[}{]}

\DeclarePairedDelimiter\bra{\langle}{\rvert}
\DeclarePairedDelimiter\ket{\lvert}{\rangle}

\newcommand{\boldtheta}{\boldsymbol\theta}
\newcommand{\boldbeta}{\boldsymbol\beta}
\newcommand{\boldgamma}{\boldsymbol\gamma}
\newcommand{\boldphi}{\boldsymbol\varphi}
\newcommand{\boldDelta}{\boldsymbol\Delta}
\newcommand{\cmax}{C_{\text{max}}}

\begin{document}

\title{Using models to improve optimizers for variational quantum algorithms}

\date{\today}
\author{Kevin J. Sung}
\email[Corresponding author: ]{kevjsung@umich.edu}
\affiliation{Google Research, Venice, CA}
\affiliation{Department of Electrical Engineering and Computer Science, University of Michigan, Ann Arbor, MI}
\author{Jiahao Yao}
\affiliation{Department of Mathematics, University of California, Berkeley, CA}
\author{Matthew P. Harrigan}
\affiliation{Google Research, Venice, CA}
\author{Nicholas C. Rubin}
\affiliation{Google Research, Venice, CA}
\author{Zhang Jiang}
\affiliation{Google Research, Venice, CA}
\author{Lin Lin}
\affiliation{Department of Mathematics, University of California, Berkeley, CA}
\affiliation{Computational Research Division, Lawrence  Berkeley National Laboratory, Berkeley, CA}
\author{Ryan Babbush}
\affiliation{Google Research, Venice, CA}
\author{Jarrod R. McClean}
\email[Corresponding author: ]{jmcclean@google.com}
\affiliation{Google Research, Venice, CA}

\begin{abstract}
Variational quantum algorithms are a leading candidate for early applications on
noisy intermediate-scale quantum computers. These algorithms depend on a
classical optimization outer-loop that minimizes
some function of a parameterized
quantum circuit. In practice, finite sampling error and gate errors make this a stochastic
optimization with unique challenges that must be addressed at the level of the optimizer. The sharp trade-off between precision and
sampling time in conjunction with experimental constraints necessitates the development of new optimization strategies to
minimize overall wall clock time in this setting.
In this work, we introduce two optimization methods and numerically compare their performance with
common methods in use today. The methods are surrogate model-based algorithms designed to
improve reuse of collected data. They do so by
utilizing a least-squares quadratic fit of sampled function values within a moving trusted region to estimate the gradient or a policy gradient.
To make fair comparisons between optimization methods, we develop experimentally relevant cost models
designed to balance efficiency in testing and accuracy with respect to cloud quantum computing systems. 
The results here underscore the need to both use relevant cost models and optimize hyperparameters of existing optimization
methods for competitive performance.
The methods introduced here have several practical advantages in realistic experimental settings, and we have used one of them successfully in a separately
published experiment on Google's Sycamore device.
\end{abstract}

\maketitle

\section{Introduction}

With recent developments in quantum hardware, including the ability to perform select tasks faster than
classical supercomputers~\cite{arute_supremacy_2019}, the push towards practical applications on these devices has intensified.
Variational quantum algorithms are among the top candidates for early applications
on noisy intermediate-scale quantum (NISQ) computers \cite{peruzzo2014variational,mcclean2016theory,Preskill2018nisq}.
These algorithms can be used to approximate ground
energies of Hamiltonians or find approximate solutions
to discrete optimization problems.
A main component of these algorithms is the minimization of
some function of
a parameterized quantum state, where that function is measured using the quantum computer.
Commonly, the function is
the expectation value of a Hamiltonian, determined by the problem of interest.
The presence of sampling error and gate errors makes the function stochastic, and
the stochasticity due to sampling error is fundamental to measuring the values on a quantum device.  
The output of this stochastic function is fed to a classical optimizer, and it is those optimizers
and constraints presented by real devices that we will focus on here.

As the classical optimizers are at the core of variational quantum algorithms, their performance
can determine the resources required to solve a problem.  Non-linear optimization of continuous
functions of the type that exist in variational quantum algorithms are commonplace in fields like
machine learning, but quantum systems offer unique trade-offs that must be considered to improve efficiency.
Given the current focus on these algorithms and the core role played by the optimizer, there have been a number of
works evaluating the performance of optimizers for different problems and contexts.
For example, at least two experimental implementations of variational algorithms
\cite{peruzzo2014variational, hempel2018quantum}
used the Nelder-Mead simplex algorithm \cite{nelder_mead_1965} to optimize the objective
function. Other experimental implementations
\cite{kandala2017hardware, otterbach_hybrid_2017, colless_computation_2018, kokail_variational_2019, pagano_qaoa_2019}
used algorithms including
Simultaneous Perturbation Stochastic Approximation (SPSA)
\cite{spall_spsa_1992},
Bayesian optimization \cite{shahriari_bayesian_2016},
particle swarm optimization
\cite{parsopoulos_pso_2002},
dividing rectangles \cite{jones_lipschitzian_1993},
and gradient descent.  In addition, there have been a number of numerical investigations of optimization
in the context of
variational quantum algorithms.
Several of these studies introduce novel heuristics and test them
numerically on example problems
\cite{wecker2015progress, zhou_qaoa_2018, nakanishi_sequential_2019, parrish_jacobi_2019, kubler_adaptive, arrasmith_frugal_2020}.
Other work \cite{guerreschi_practical_2017, romero2017strategies, nannicini_hybrid_2019, yao_policy_2020, lavrijsen_optimizers_2020,leng_robust_2019, stokes_natural_2020, wierichs_avoiding_2020} has compared the performance of methods including Nelder-Mead,
limited-memory Broyden-Fletcher-Goldfarb-Shanno, \cite{byrd_lbfgsb_1995}, Constrained Optimization
By Linear Approximation \cite{powell_cobyla_1994},
Powell's method \cite{powell_efficient_1964},
SPSA, RBFOpt \cite{costa_rbfopt_2018}, Stable Noisy Optimization
by Branch and Fit \cite{snobfit},
Bound Optimization by Quadratic Approximation \cite{bobyqa},
Mesh Adaptive Direct Search \cite{nomad},
implicit filtering \cite{imfil}, policy-gradient-based reinforcement learning \cite{williams_connectionist_1992}, and natural gradient \cite{stokes_natural_2020}.

There is a considerable body of work in evaluating optimizers for use in variational algorithms, but not all of these works
use cost metrics relevant to quantum experiments.  For example, it is common to evaluate a suite of optimizers
based on number of optimizer iterations required for convergence to a local optima, using noiseless function
evaluations.  However, the inherent quantum nature of the sampling procedure implies that the first iteration could
have taken an unbounded amount of experimental time in such a setup (noiseless evaluation), 
and hence conclusions based on such studies may not be applicable to experiments.
A meaningful comparison of these methods must treat the stochastic nature of the objective
function and related costs in terms of experimental time to solution to properly compare methods.
While some past works do account for the effect of stochastic noise
\cite{kubler_adaptive, arrasmith_frugal_2020, yao_policy_2020, lavrijsen_optimizers_2020},
in this work we additionally incorporate other experimental parameters
into our cost models.
In developing our models, we focus on the case of superconducting quantum computers accessed 
through the Internet, though our models can be easily modified for other architectures.
We account for parameters such as the sampling rate of the quantum processor
and the latency
induced by communicating over the Internet.
The proper choice of optimizer ultimately depends on the details of the experiment constraints.

In consideration of constraints we did not find satisfied in other methods, 
we introduce two surrogate model-based optimization algorithms we call Model Gradient Descent (MGD) and Model Policy Gradient (MPG)
and numerically compare their performance against 
commonly used methods. In particular, we target the tendency for
local methods to
under-utilize the existing history of function evaluations.  We have successfully used MGD
in an experimental implementation of the Quantum Approximate
Optimization Algorithm \cite{farhi_qaoa1_2014} on a
superconducting qubit processor \cite{arute_qaoa_2020}.
We perform systematic tuning of optimizer hyperparameters
before comparison for all methods, and measure performance using
estimates of actual wall clock time needed in a
realistic experimental setting.  An important, though unsurprising, implication of our results is that
hyperparameter tuning under the correct cost models is crucial for performance in practice.

The outline of this work is as follows.  In \sec{problems_and_cost_models} we set up the example problems we study and
describe in more depth the problem of developing efficient cost models to allow comparison
of methods.  In \sec{optimizers} we describe the optimizers we study and how we tuned their hyperparameters. After this setup, we compare the performance of optimizers numerically
in \sec{results} using our developed cost models.  At a glance, our results highlight the importance
of different cost model features, how constraints influence the optimal choice of optimizer, and 
the importance of hyperparameter optimization.  Stochastic optimizers with hyperparameters permitting
varying levels of noise in the objective are found to be generally more robust and efficient.  Finally,
we end with some concluding thoughts in \sec{conclusion}.

\section{Problems studied and cost models}
\label{sec:problems_and_cost_models}

\subsection{Problems studied}
As the performance of an optimizer can be intimately tied to the problem studied, it is important to look at a range of problems
in evaluating their relative performance. As two of the most common areas studied in variational quantum algorithms are combinatorial
optimization and ground state preparation of fermionic systems, we select these for our sample problems.  Here we
aim to clarify the details of the systems, circuit ansatze, and initial parameters modeled in our numerical tests.

While multi-modality of cost functions is an important consideration in variational quantum algorithms,
it turns out that even optimization within a single convex basin can be challenging enough to warrant independent
investigation due to constraints imposed by the quantum device.  To this end, we assume throughout that we have knowledge of an
initial guess which is in the convex vicinity of an optimum and our goal
is simply to converge to that local optimum. Several strategies have been proposed
for choosing such an initial guess
in contexts including optimization and chemistry \cite{zhou_qaoa_2018, brandao_qaoa_2018, romero2017strategies, wecker2015progress}.

\subsubsection{Max-Cut on 3-regular graphs}

The maximum cut problem (Max-Cut) is widely studied and known to be NP-hard. It has been used in several
previous experimental implementations of
variational quantum algorithms
\cite{otterbach_hybrid_2017, arute_qaoa_2020} and hence allows for straightforward performance comparisons.
The problem is specified by an undirected graph
on $n$ vertices and the goal is to label each vertex with either
$+1$ or $-1$ in order to maximize the number of edges whose vertices
have different labels. This cost function is represented 
by the Hamiltonian
\begin{align}
    C = \sum_{\langle i, j \rangle} \frac12 (I - Z_i Z_j),
\end{align}
where $Z_j$ is the standard Pauli $Z$ operator applied to qubit $j$ which is node $j$ on the graph, and $\langle i, j \rangle$ ranges over the edges of the graph. The goal is to
find a computational basis state that \emph{maximizes} the Hamiltonian.

We use the Quantum Approximate Optimization Algorithm (QAOA) \cite{farhi_qaoa1_2014}
ansatz used to approximately solve the Max-Cut problem on random 3-regular graphs.
The QAOA ansatz depends on the number of rounds, $p > 0$, and is parameterized
by $2p$ real numbers $\boldgamma = (\gamma_1, \dots, \gamma_p)$ and $\boldbeta = (\beta_1, \dots, \beta_p)$.
The ansatz is
\begin{align}
    \ket{\boldgamma, \boldbeta} = U_B(\beta_p)  U_C(\gamma_p ) \cdots U_B(\beta_1) U_C(\gamma_1 )  \ket{+}^{\otimes n},
\end{align}
where
\begin{align}
    U_C(\gamma) = e^{-i \gamma C}, \quad U_B(\beta) = e^{-i \beta B}, \quad B = \sum_{i=1}^n X_i,
\end{align}
and $\ket{+}^{\otimes n}$ is the uniform superposition of all $2^n$ computational basis states.

For our numerics, we focus on a randomly chosen instance to minimize the number of uncontrolled variables. 
Moreover, for QAOA focusing on a single instance is justified because the optimization landscape has been shown to 
concentrate for different randomly chosen instances~\cite{brandao_qaoa_2018}.
To obtain an initial guess for this problem, we classically
computed a locally optimal parameter vector and then perturbed it with a uniformly random vector of length $0.1$.
At $p=1$ the optimal parameter vector had a length of 0.462,
and at $p=5$, 1.285.

In our numerics we report the approximation ratio
\begin{align}
    \frac{\bra{\boldgamma, \boldbeta} C \ket{\boldgamma, \boldbeta} }{\cmax}
\end{align}
where $\cmax = \max_{z} \bra{z} C \ket{z}$.
The goal is to maximize this value, which falls in the range $[0, 1]$.

\subsubsection{Sherrington-Kirpatrick model}

Another model we consider is the Sherrington-Kirkpatrick (SK) model \cite{sherrington_kirkpatrick_model_1975},
which is a canonical example of a frustrated spin glass.
It has been used in at least one previous
experimental implementation of variational
algorithms \cite{arute_qaoa_2020}.
The Hamiltonian is given by
\begin{align}
    H = \sum_{i < j} J_{ij} Z_i Z_j
\end{align}
where $J_{ij}$ is selected uniformly at random from $\set{-1, 1}$.
We use the QAOA ansatz to approximate the solution of this problem, by minimizing
the expected cost.

Again, for our numerics we focus on a single randomly generated instance, where generality of performance
is supported by concentration results in QAOA.
As an initial guess for this problem, we classically
computed a locally optimal parameter vector and then perturbed it with a uniformly random vector of length $0.1$.
At $p=1$ the optimal parameter vector had a length of 0.452,
and at $p=5$, 1.044.

For comparison between problems, we normalize energy values $E$ to new values $E'$
by the formula
\begin{align}
    E' = \frac{E - E_{max}}{E_{min} - E_{max}}
\end{align}
where $E_{min}$ and $E_{max}$ are the lowest and highest eigenvalues
of the Hamiltonian, respectively. Thus we are in fact maximizing this
normalized energy value, which falls in the range $[0, 1]$.

\subsubsection{Hubbard model}

We study the task of approximating the ground state energy of the 2-dimensional Hubbard model
\cite{hubbard_model_1963}, a widely studied model that
has resisted exact solution for decades in large size limits. It is believed to be relevant to understanding
high-temperature superconductivity \cite{dagotto_correlated_1994}.
The Hamiltonian of the Hubbard model is
\begin{align}
    H
    &= -t \sum_{\langle i, j \rangle, \sigma} (a_{i, \sigma}^\dagger a_{j, \sigma} + a_{j, \sigma}^\dagger a_{i, \sigma})
    \notag \\
    & + U \sum_{i} a_{i, \uparrow}^\dagger a_{i, \uparrow} a_{i, \downarrow}^\dagger a_{i, \downarrow}
    \label{eq:hubbard_hamiltonian}\\
    &= T + V \\
    &= T_h + T_v + V
\end{align}
where the $a_{i, \sigma}$ are fermionic annihilation operators,
$\langle i, j \rangle$ ranges over edges in the lattice,
$\sigma \in \{\uparrow, \downarrow\}$ is a spin degree of freedom,
and we have split the sum into the hopping term $T$ and interaction term $V$.
$T$ is further decomposed into sub-terms $T_h$ and $T_v$
corresponding to horizontal and vertical edges, respectively.
We set $t=1$ and $U = 4$ for our numerical experiments, which corresponds to
a regime of modest correlation ill-suited for mean-field methods.

We use a ``Hamiltonian variational'' ansatz similar to the one in ref.~\cite{wecker2015progress}.
It is inspired by the idea of state preparation via adiabatic evolution.
Similar to QAOA, our ansatz has a basic circuit repeated $p$ times, but for flexibility it is
varied non-uniformly with respect to hopping. 
The basic circuit has three parameters which we call $\theta_h$, $\theta_v$, and $\theta_U$, and it approximates a unitary of the form
\begin{align}
    \exp\bracks{-i (\theta_h T_h + \theta_v T_v + \theta_U V)}
\end{align}
The approximation is achieved using a second-order Trotter step based on the fermionic swap network \cite{kivlichan2018quantum},
in which a swap network is used to apply the terms of the
Hamiltonian and then the same network is applied but in
reverse order. Because the swap network can be implemented with only
linear qubit connectivity, this ansatz is amenable to implementation
on near-term superconducting qubit hardware.
The ansatz is similar to the one used in ref.~\cite{wecker2015progress} but corresponds to a different ordering of terms.
In total there are $3p$ parameters.

We study the model at half-filling. Our numerics are performed on the $2 \times 2$ system, which under
standard encodings corresponds to an 8 qubit system.
For our initial state we use a ground state of the hopping term that is
precisely described in Appendix \ref{app:initial_state}. This state is easy to prepare
on a quantum computer and is expected to be adiabatically connected to the ground state of $H$
for modest values of $t/U$.
For our initial guess, we set the parameters so that the ansatz circuit consists of a sequence
of second-order Trotter steps approximating the dynamics of the
time-dependent Hamiltonian $H(t) = T + (t/A)V$ for $t \in [0, A]$, where
$A = 0.1\cdot Up$.
This choice is motivated by the idea of state preparation via adiabatic evolution.

As with the Sherrington-Kirkpatrick model, we normalize energy values $E$ to new values $E'$
by the formula
\begin{align}
    E' = \frac{E - E_{max}}{E_{min} - E_{max}}
\end{align}
where $E_{min}$ and $E_{max}$ are the lowest and highest eigenvalues
of the Hamiltonian, respectively. Thus we are in fact maximizing this
normalized energy value, which falls in the range $[0, 1]$.

\subsection{Cost models}
\label{sec:cost_model}

An essential element of developing and improving optimizers for variational algorithms is an accurate
cost model that respects the quantum nature of the problem and imperfections of the device.  Studies
that restrict evaluation of optimizers to abstract ``number of iterations'' using perfect function
queries can yield faulty conclusions and hide the implication that a single function evaluation to that precision
could have taken years or more.  A core challenge is the stochastic nature of the function evaluation and
shot limited precision in the estimates.  Moreover, imperfections in the device and implementation can complicate
matters.  Unfortunately, without a quantum device, precise simulation of the impact of noise can be prohibitively
expensive, and so a balance must be struck between accuracy and cost effectiveness of the simulations to
maximize applicability.  Here we detail how we construct our models to strike this balance.

We restrict our interest to minimizing the expected energy of a Hamiltonian $H$ with efficient
Pauli expansions $H = \sum_j \alpha_j P_j$ (in the case of the Hubbard model (\ref{eq:hubbard_hamiltonian}),
the Jordan-Wigner Transformation \cite{jordan1928uber} is applied to obtain the Pauli expansion), so the objective function is
\begin{align}
    f(\boldtheta) = \langle \boldtheta \rvert H \lvert \boldtheta \rangle,
\end{align}
where $\lvert \boldtheta \rangle$ represents the
ansatz state with parameters $\boldtheta$.
Most of the optimizers that we present results for use queries to the objective function without
any additional kinds of queries, but we also present results for stochastic
gradient descent, which queries the gradient.

\subsubsection{Objective function queries}
The exact estimator used to query the objective function on the quantum device can
take a wide variety of forms depending on factors in the device and the problem of interest.
At a glance, however, a query to the objective function is often answered by measuring the expectation values
of the terms $P_j$ and using the coefficients $\alpha_j$ to form an estimate of $f(\boldtheta)$.
When simulated in the most accurate way, the measurement of each individual term implies
a variance on the estimate which is state-dependent, and functions like a Bernoulli random
variable.  Moreover, the variance of that measurement can be influenced by parallel measurements
being performed, even when they commute~\cite{mcclean2016theory}.  Trade-offs in the influences of these factors
have inspired recent research in developing more efficient estimators with a given number
of samples~\cite{jena_pauli_2019, izmaylov_revising_2019, huggins2019efficient, izmaylov_unitary_2020, verteletskyi_measurement_2020}.  However, perfect emulation of these proposals can be prohibitively expensive, even
in classical simulation of small systems, and hence it is desirable to develop models of the
process that strike a good balance between accuracy and simulation cost so that the full
variational process can be simulated on a range of systems.

In the cases of Max-Cut and the Sherrington-Kirkpatrick model, the Hamiltonian is diagonal
and all of its terms can be measured simultaneously in one shot. In our numerical experiments,
we simulated these measurements directly.  However, for non-diagonal Hamiltonians such as the
Hubbard model, we take a different strategy.

As there are many terms in the sum, which are typically evaluated by repeated and independent measurement,
a Gaussian random function query turns out to be a good and extremely cost effective model.
That is, in our simulations a query to the objective function is modeled as
\begin{align}
    f(\boldtheta) = \langle \boldtheta \rvert H \lvert \boldtheta \rangle + \mathcal{N}(0, \lambda^2/M)
\end{align}
$\langle \boldtheta \rvert H \lvert \boldtheta \rangle$ is evaluated exactly,
$\mathcal{N}(\mu, \sigma^2)$ is a normal random variable with mean $\mu$ and variance $\sigma^2$, 
and $M$ is the number of repeated experiment repetitions.
Note that even in the presence of hardware errors, the expectation value of the Hamiltonian
would be the sum of many independent random variables, so this would still be a good model.
Here, we estimate the variance is using
a known lower bound for common measurement strategies,
previously derived for the general case
\begin{align}
    \lambda^2 = (\sum_j |\alpha_j|)^2
\end{align}
which empirically we have observed to be loose when compared with exact models, but qualitatively
matches the behavior and overestimates the number of measurements by a factor of 2 in many cases.
We note that a wealth of other strategies have been developed to shrink the effective variance for a fixed
number of queries $M$~\cite{jena_pauli_2019, izmaylov_revising_2019, huggins2019efficient, izmaylov_unitary_2020, verteletskyi_measurement_2020}, but we do not consider them in detail here.
Since the bound we use is a worst-case bound that is independent of the quantum state,
our cost estimates are likely to be conservative.

The dependence of the variance of the
estimate on the number of samples represents a key trade-off we consider in many algorithms here, as some
optimizers can tolerate heavier amounts of noise than others, and hence we take the number of shots at each
iterate to be an important hyperparameter.
In our numerical experiments on the Hubbard model, we simulated queries by computing
the exact expectation value and then artificially adding noise drawn from a normal distribution, using this bound to determine the variance of
the distribution for a specified number of measurement shots.

\subsubsection{Gradient queries}
For optimizers that use analytic gradient queries, we assume that queries to the gradient of the objective function are answered
by applying the ``parameter-shift rule''
\cite{mitarai_learning_2018, schuld_gradients_2019, crooks_parameter_2019}.
This is a method of obtaining an unbiased estimator of the gradient without
using ancilla qubits, and applies to ansatze of the form
\begin{align}
    \lvert \boldtheta \rangle = \exp(-i \theta_p A_p) \cdots \exp(-i \theta_1 A_1) \lvert \psi \rangle
\end{align}
where for our purposes each $A_j$ is a Hermitian sum of commuting Pauli matrices.
The technique exploits the fact that if $A_j$
has two eigenvalues $\pm r$, then
$\frac{\partial f}{\partial \theta_j} (\boldtheta) = r(f(\boldtheta^+) - f(\boldtheta^-))$
where $\boldtheta^+$ is $\boldtheta$ but with the $j$-th coordinate equal to $\theta_j + \frac{\pi}{4r}$
and $\boldtheta^-$ is $\boldtheta$ but with the $j$-th coordinate equal to $\theta_j - \frac{\pi}{4r}$.
If some parameters are constrained
to be the same, then the derivative is obtained by summing the results
of this expression for each parameter; the number of objective function queries needed is then
two times the number of those parameters.
If $A_j = \sum_k P_k$ for commuting Pauli operators $P_k$, then we decompose
$\exp(-i \theta_j A_j) = \prod_k \exp(-i \theta_j P_k)$ and then apply the previous rule.
Thus, the cost of evaluating the partial derivative is proportional to the number of terms in the sum,
in a loose way.  In practice, this sum is evaluated stochastically with a probability depending on
the weight of the term in the sum~\cite{Harrow2019low}.

\subsubsection{Wall clock time}

Ultimately, one is interested in minimizing the amount of time it takes to run a complete experiment
to some fixed precision.  The models we develop here are meant to capture this in a cost efficient
way, without using a wildly inaccurate proxy like mere ``number of optimizer iterations''.  To this end,
we not only consider the sampling noise, but also constraints like latency concerns inherent to real experiments.

To estimate the running time of an experiment we develop a model based on superconducting
qubits \cite{barends_surface_2014, corcoles2015demonstration}. We also assume
the user is executing the experiment through a cloud computing service, potentially introducing
network latency. We consider three scenarios regarding network latency: zero latency,
corresponding to the optimizer running completely on the server side; circuit batching,
in which the user is allowed to send multiple circuits to the service in one batch; and finally
no circuit batching, where the user is only allowed to send one circuit at a time.

The total running time of an experiment is equal to the number of queries made times
the amount of time it takes to satisfy a single query. The time needed to satisfy
a single query can be split into the time $T_\text{sample}$ used in sampling
circuits on the quantum processor, the time $T_\text{switch}$ representing
the overhead in switching between different circuits, and $T_\text{cloud}$
representing the latency in communicating over the Internet.
We have $T_\text{sample} = M/s$ where $M$ is the number of measurements
made to satisfy the query and $s$ is the sampling rate of the processor;
$T_\text{switch} = r \times c$ where $r$ is the overhead in readying
the quantum processor to execute a circuit and $c$ is the number of different
circuits executed; and $T_\text{cloud} = \ell \times c / b$ where
$\ell$ is the network round-trip time for communicating with the cloud server
and $b$ is the number of circuits sent to the server in a single round
of communication.
We use the values $s = 10^5$ Hz and $r = 0.1$ s. This sampling rate has not yet been
achieved experimentally but is plausible assuming an order of magnitude or two
improvement in current capabilities is possible; a recent experiment achieved
a sampling rate of about $5 \times 10^3$ Hz \cite{arute_supremacy_2019}.
When including network latency, we set $\ell = 4.0$ s; this value is based on our
own experience executing experiments through an internal cloud interface.
The  value of $b$ depends on the details of the algorithm.
We ignore as negligible the time taken by the classical optimization algorithm to select parameters
for querying, as the optimizers here use relatively simple classical updates.

\section{Optimization strategies}
\label{sec:optimizers}

\subsection{Choice of optimizers}

A wide range of optimizers now exist for continuous, non-linear optimizations, with different
strengths and weaknesses.  One key element for consideration is the stochastic nature of our
objective function and its relation to the number of measurements made for each function evaluation.  Some
optimizers were designed with noiseless (up to reasonable precision limits) function evaluations
in mind, and are relatively unstable with respect to even small amounts of noise.   While one could 
insist on a number of measurements that renders the function evaluations essentially exact, 
this incurs a huge overhead per iteration.  We group algorithms into two categories, distinguished
by whether they have inherent hyperparameters that allow them to adjust their resilience to noise.
If an algorithm in practice requires that the input be given to a fixed precision in order to be stable,
we term it deterministic.  If it has a hyperparameter that naturally allows it to accept more or less noise,
we call it stochastic.

The difference between the two classes can be subtle, and depend on the details of implementation.
For example, a gradient descent implementation that makes use of an exact line search can accidentally
rule out good regions of space from small wobbles in a query value, and is hence deterministic.  However,
if that sample implementation substitutes a fixed step with a learning rate, it is not only more robust
to noise, but that learning rate can be adjusted to match noise levels in the objective queries.  Hence we term that 
a stochastic optimizer.  Considering the costs of each with external hyperparameters (e.g. number of measurements)
and internal hyperparameters (e.g. learning rate) tuned for optimal performance will show us these trade-offs.

Overall, we investigated six different optimizers. Four of these have been
studied in past work, and the last two are surrogate model-based
optimizers that we introduce here.
Surrogate model-based optimizers construct a model of the
objective function using previously evaluated points and use the model
to determine what points to evaluate next. They are popular choices for the
optimization of objective functions that are expensive to evaluate or noisy (or both)
\cite{jones_taxonomy_2001, cartis_improving_2019}.

Listed briefly, the optimizers we study here are:
\begin{itemize}
    \item Deterministic algorithms:
    \begin{itemize}
        \item The Nelder-Mead simplex method \cite{nelder_mead_1965}. This method has been used
        in previous theoretical \cite{guerreschi_practical_2017, romero2017strategies} and experimental
        \cite{peruzzo2014variational, hempel2018quantum} works on variational algorithms.
        We used the implementation from SciPy \cite{scipy_2020}.
        \item Bounded Optimization By Quadratic Approximation (BOBYQA) \cite{bobyqa}.
        This is a surrogate model based algorithm that uses an interpolating quadratic
        model to approximate the objective function, and has been studied
        in a previous work on variational algorithms \cite{lavrijsen_optimizers_2020}.
        We used the implementation from the Python package Py-BOBYQA \cite{cartis_improving_2019}.
    \end{itemize}
    \item Stochastic algorithms:
    \begin{itemize}
        \item Simultaneous Perturbation Stochastic Approximation (SPSA) \cite{spall_spsa_1992}.
        This method has also been used in previous theoretical \cite{nannicini_hybrid_2019}
        and experimental \cite{kandala2017hardware} works on variational algorithms.
        We used our own implementation.
        \item Stochastic gradient descent using analytic gradient measurements obtained via
        the ``parameter-shift rule''
        \cite{mitarai_learning_2018, schuld_gradients_2019, crooks_parameter_2019}.
        \item Model Gradient Descent (MGD).
        This is a surrogate model-based algorithm we introduce here that uses a least-squares quadratic model to
        estimate the gradient of the objective function. We give pseudocode in Appendix \ref{app:pseudocode}.
        \item Model Policy Gradient (MPG).
        Building on the vanilla policy gradient method\cite{yao_policy_2020}, this method additionally introduces a least-squares quadratic model to reduce the variance in the estimation of the policy gradient. We give pseudocode in Appendix \ref{app:pseudocode}.
    \end{itemize}
\end{itemize}

\subsection{Model gradient descent and policy gradient}

In this section we describe and motivate the design choices of our new algorithms, Model Gradient Descent and Model Policy Gradient,
which are described in pseudocode in Algorithm \ref{alg:mgd} and \ref{alg:pg}. These are surrogate model-based methods
which use least-squares regression to fit quadratic models of the objective function.  A key
expense in variational quantum algorithms is the evaluation of the function at different points, which is costly due to the underlying variance.  Hence, it would be beneficial to reuse the
history of point evaluations, rather than to discard them at each iteration. For local optimizations
where iterates proceed gradually, it seems intuitive that this should be possible.  Eventually,
if one collected enough points in a small enough region, it should be possible to construct a surrogate
model that is more accurate than raw function evaluations at a fixed number of measurements.

As a combination of this motivation and simplicity, we use a least-squares fit to a quadratic function. 
However, it is also clear that if the region of sampled points is too large, the function may not be well
approximated by a quadratic, hence we use a trusted region of sample points, which may be new or reused from
previous iterates.

In each iteration, the algorithms sample a number of points randomly from the vicinity of the current iterate.
They fit quadratic models to these points and other previously evaluated points within the vicinity.
Finally, MGD uses the gradient of this quadratic model as an approximation to the true gradient
and performs gradient descent; MPG queries the model to evaluate a large batch of data points and performs policy gradient optimization.
The reason we
did not use standard trust-region solution techniques after building
the quadratic model is that we found empirically that the eigenvalues of the Hessian of the quadratic model  built
upon stochastic function evaluations may be slightly negative, which dictates in a standard
trust region solution method that the solution is on the exterior of the trust region.  This constant jumping
to the exterior of the trust region represented a sort of fundamental inefficiency under stochastic functions.
In contrast, the gradient or policy gradient of the model, while stochastic, represented a reliable estimator that, in conjunction
with techniques like a fixed learning rate, combined the increased accuracy of additional samples with the
robustness of a stochastic gradient descent.

To enhance the performance and stability of the methods, 
we introduced several hyperparameters to our algorithms.
In particular, as algorithms approach an optimum, decreasing the radius of the neighborhood
from which points are sampled is expected to give a more accurate estimate of the function value and its gradient.
Thus, we introduce a hyperparameter $\xi$ for MGD which controls the rate at which the radius decreases. As for MPG, we introduce the fixed sample radius ratio $\delta_r$ with respect to the maximal sample radius of the policy. The selected sample radius adaptively shrinks along with the maximal sample radius as the policy gradually becomes more confident.
It may also be advantageous to decrease the learning rate of both algorithms.
Thus, we introduce hyperparameters $\alpha$ and $A$ which control the rate of this decrease.
The parameters $\xi$ and $\alpha$ are exponents for geometric decay, which is a standard
way to scale parameters like learning rates throughout an optimization algorithm, used in methods such as SPSA.
The details
of how these parameters enter can be found in the pseudocode of the algorithms.

\subsection{Hyperparameter selection}

\begin{figure}
    \centering
    \includegraphics[width=1.0\linewidth]{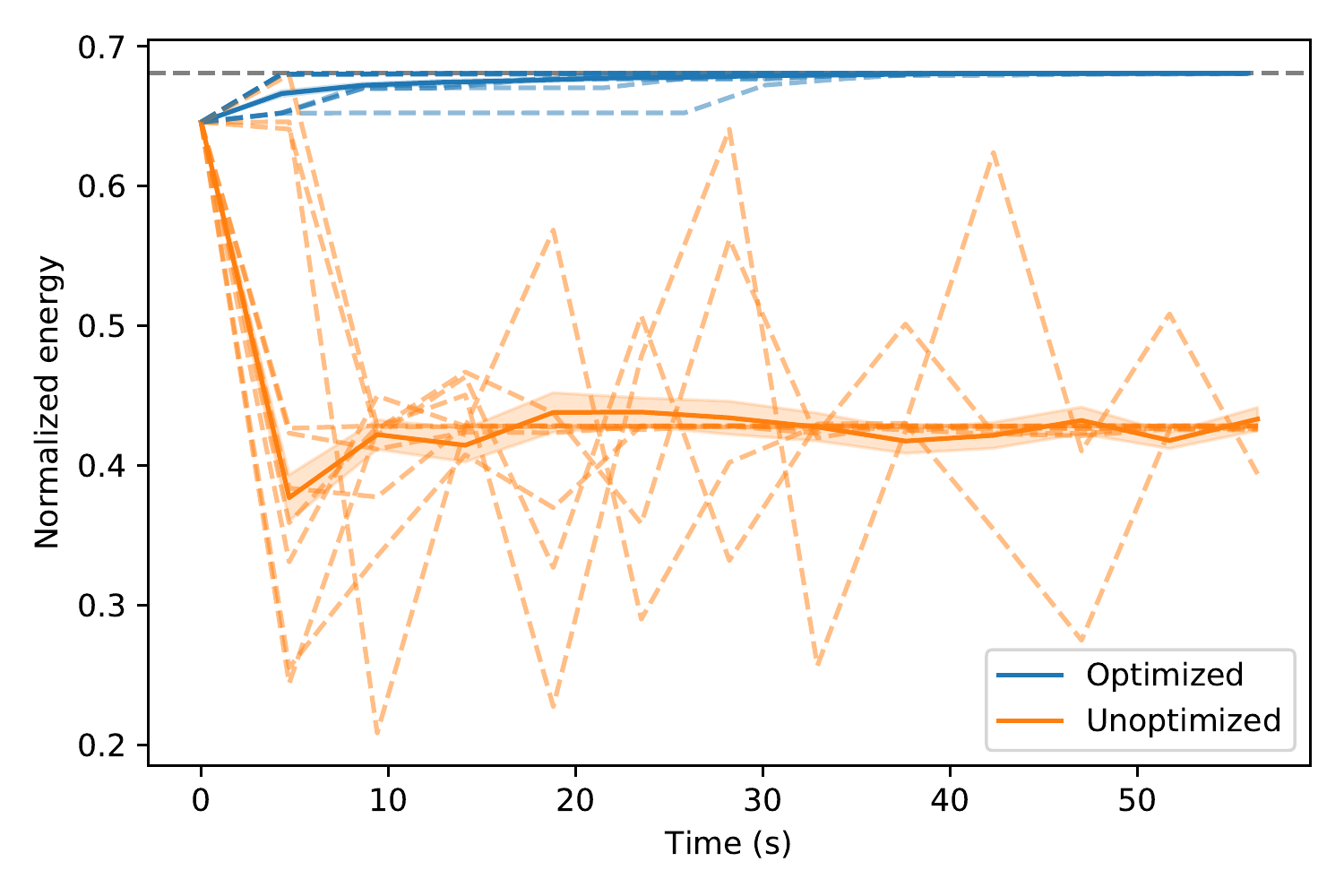}
    \caption{Optimization progress of SPSA in simulated experiments on a Sherrington-Kirkpatrick
    model Hamiltonian using two different hyperparameter settings:
    the ones used by default in the implementation from the software package Qiskit
    (Unoptimized), and ones that were found by searching for good settings (Optimized).
    The solid line represents the mean energy over 50 runs
    with different PRNG seeds, and the shaded region represents
    a width of one standard deviation of the mean. The dotted lines are
    10 example trajectories.
    The dotted gray line corresponds to the ansatz optimum. SPSA
    fails to converge with the unoptimized hyperparameters.}
    \label{fig:hyperparameters}
\end{figure}

Each optimizer we considered here has a number of hyperparameters, and empirically we noted that 
the choice of these hyperparameters had a great impact on performance. Strikingly, some optimizers that failed
completely with out of the box settings became competitive choices with even slight adjustments.  
Recalling that many of the optimizers we consider are inherently deterministic, one important hyperparameter external to all methods 
is the number of measurement shots per energy evaluation.

We tuned hyperparameters by grid search, and
separately for each problem class and ansatz depth considered. For each combination of hyperparameters
considered in the search, we performed an optimization run using the wall clock time model that includes 
network latency and circuit batching.
The optimal hyperparameters were those that minimized time to convergence with a precision target of $10^{-3}$. 
To avoid effects of overfitting, we restricted consideration to single realizations, where other runs
are not further optimized within a problem class.
Note that the details of hyperparameter selection has a significant effect on the
performance of the algorithms.  For example, choosing a more lenient precision requirement while still minimizing time
to solution leads to different performance characteristics on other problems.
See
Appendix \ref{app:hyperparameters}
for more details, including descriptions of the hyperparameters.

As a simple demonstration of the importance of hyperparameter selection,
we considered the performance on a simple
test case with two different hyperparameter settings.
\fig{hyperparameters} shows the optimization progress of SPSA in
simulated experiments on a Sherrington-Kirkpatrick model Hamiltonian with
$n=8$ and $p=1$, using two different hyperparameter settings:
the ones used by default in the implementation from the software package Qiskit \cite{Qiskit},
and ones that we optimized for minimal time to solution with a fixed precision cutoff.
Depicted is the normalized energy versus wall clock time,
using the wall clock time model that includes network latency and circuit batching.
With tuned hyperparameters,
SPSA converges to the solution rapidly, and without tuning it quite obviously does not.
The erratic trajectory when using the unoptimized default
parameters can be attributed to the fact that the initial
learning rate of the algorithm is set to a value
over 100 times larger than the optimized value.
Hence, while SPSA is a powerful stochastic
method capable of dealing with variable function noise,
hyperparameter tuning must be actively used to make a proper comparison.
Not taking advantage of this capability has led previous studies
to underestimate the performance of SPSA or outright conclude that
it is not effective for these problems
\cite{nakanishi_sequential_2019, nannicini_hybrid_2019}.
This demonstrates the importance of tuning hyperparameters in making a fair comparison
between optimization algorithms, and throughout this study we tune all methods under consideration.

\section{Results}
\label{sec:results}

\begin{figure}
\includegraphics[width=\linewidth]{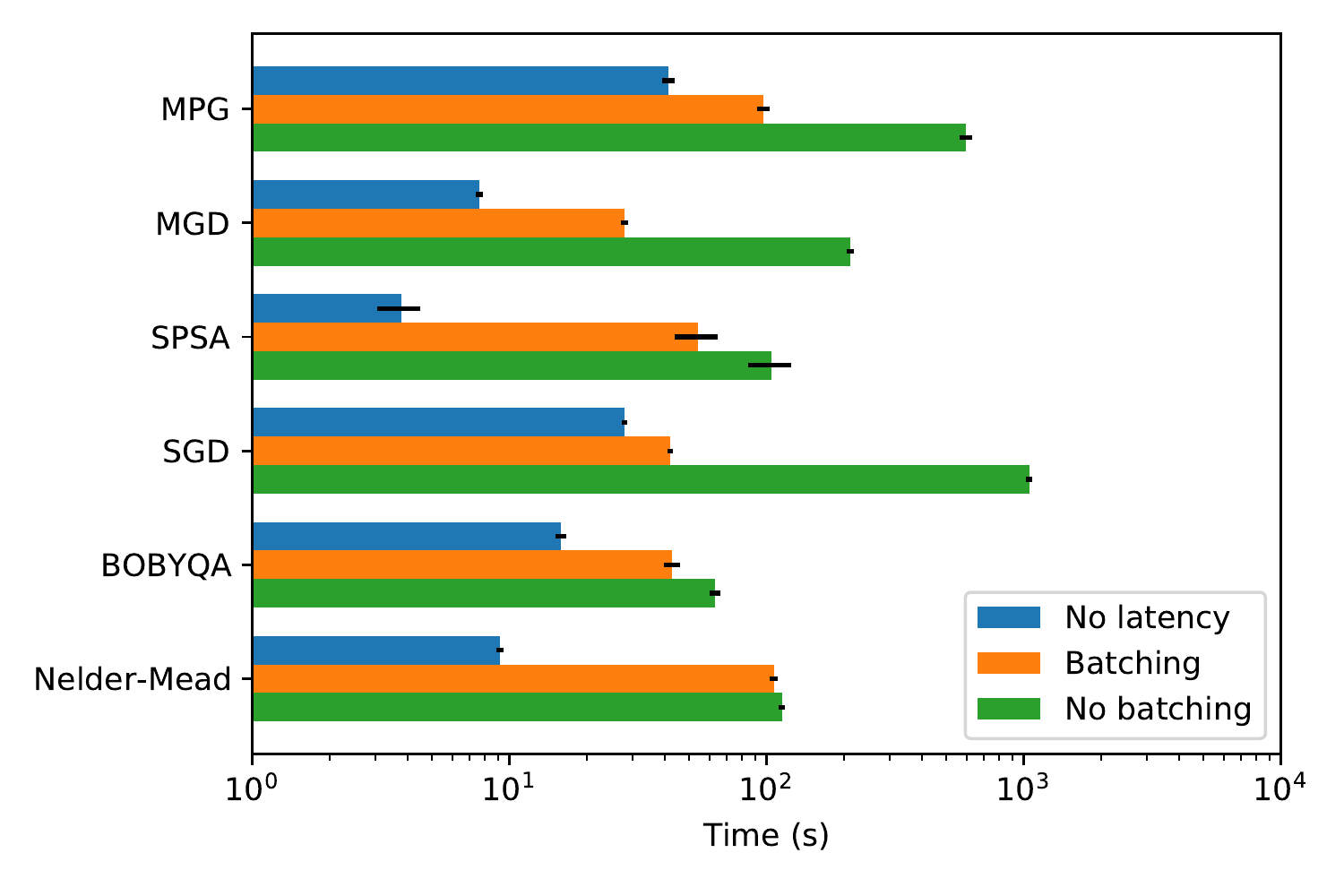}
\caption{Wall clock time for optimization to achieve precision 1e-3 for the
Sherrington-Kirkpatrick model at $p=1$.
Times are averaged over 50 experiments with different PRNG seeds. The black lines at the tips of the bars
represent a width of one standard deviation.
The best choice of optimizer can depend on the wall clock time model, with MGD, MPG, and SGD benefiting greatly from the ability to request execution of a batch of circuits.}
\label{fig:time_to_solution_sk_n8_p1_NO_NOISE}
\end{figure}

To increase the applicability of our results to experiment, we consider both ideal and faulty operation of
a quantum device.  In the first case, in order to isolate challenges pertaining only to sampling noise,
we assume an ideally functioning
quantum computer, so that the only source of stochasticity in the objective function is
finite sampling effects. In the other case, we modeled the effect of gate rotation error
as follows: each time the
optimizer queries the point $\boldtheta$, the objective function is evaluated
at the point $\boldtheta + \boldsymbol\varepsilon$ instead, where each component
of $\boldsymbol\varepsilon$ is chosen from the normal distribution with mean 0 and
standard deviation $\varepsilon$ (for some gate error level $\varepsilon$).
Since this error model does not straightforwardly
translate to the calculation of gradients for SGD, we did not perform
simulations of gate error with SGD.
This error model is a simplified model of coherent control error, an important
source of errors on actual hardware \cite{arute_supremacy_2019}, and which is
especially pertinent to the case of quantum computers accessed through
cloud services which are used often but calibrated only periodically.

Each simulation we perform is characterized by
four attributes: the problem (3-regular Max-Cut, Sherrington-Kirkpatrick, or Hubbard),
the ansatz depth $p$, the choice of optimizer,
and gate error level $\varepsilon$ (possibly 0).
For each set of attributes considered, we performed 50 statistically independent simulations.
For each numerical simulation we performed, we estimate the wall-clock time
of actually performing the experiment on a quantum computer accessed through a cloud service
using the various cost models described in \sec{cost_model}, and set a limit to the total
amount of time allowed. We are interested in how quickly a given optimization
algorithm converges to the optimal energy to within a target precision.
By ``optimal energy'' we mean the energy of the ansatz state at the nearest
local optimum as determined from a classical optimization of the
noiseless objective function.

\subsection{The case of \texorpdfstring{$p=1$}{p=1} and no gate errors}

First, we present the results of simulations with $p=1$ and no gate errors.
\fig{time_to_solution_sk_n8_p1_NO_NOISE} shows the wall clock time
for different optimizers to achieve precision $10^{-3}$ for the Sherrington-Kirkpatrick model at $n=8$ and $p=1$.
We define this time to be the earliest time at which
the current and all future evaluated points have an approximation ratio
or normalized energy close
to the optimal value to within $10^{-3}$.
We show the results for the three different wall-clock models described
in \sec{cost_model}: no network latency,
network latency present but with circuit batching, and network latency present
with no circuit batching. Note that Nelder-Mead converged in only 44 out of
50 runs; the other algorithms converged in all of them.

These results show that the proper choice of optimizer depends on the situation.
SPSA performed the best under the wall clock time model with
no latency, but was outperformed by MGD, SGD, and BOBYQA
under the model that included latency and circuit batching.
Under the model that included latency but did
not have circuit batching, BOBYQA performed
the  best.

The importance of the wall clock time model, and in particular the effect
of network latency, is evident. In the presence of network latency, MPG, MGD and SGD benefit much more
from circuit batching than the other algorithms do.
Both algorithms work by obtaining an estimate of the objective function gradient in each iteration.
Circuit batching provides
a benefit because multiple different circuits are needed to estimate the gradient,
and these circuits can be sent over the network
in one batch, reducing total network latency costs. SPSA also estimates the gradient, but it only uses 2 different circuits for that purpose.
In contrast, the hyperparameters of MGD and MPG were chosen so that they
both used 10, while SGD used 72. Indeed, the plot shows SGD benefiting
from batching to a greater degree than MGD.

As an illustration of the ability of the various optimizers to tolerate different amounts
of variance in the objective function, we note that the optimal hyperparameters dictates that
SGD uses 1,000 measurement shots per evaluations,
MGD and SPSA use 5,000, MPG uses 20,000, Nelder-Mead uses
25,000, and BOBYQA uses 125,000.  This makes clear our distinction between deterministic
and stochastic optimizers.  While one can find external hyperparameter settings that allow Nelder-Mead and BOBYQA
to succeed, the lack of internal hyperparameters for noise tolerance means the number of measurements grows
wildly.  In contrast, stochastic methods like MPG, MGD and SPSA can find balanced settings using far fewer measurements
per point while remaining stable.  In larger systems beyond the scope of simulation, it may not be easy to a priori 
determine the required measurements to make a deterministic method stable, and hence the flexibility of
naturally stochastic methods is likely to be preferred.  For all cases, however, some amount of hyperparameter
tuning is a necessity for good performance.

\begin{figure*}
\includegraphics[width=\linewidth]{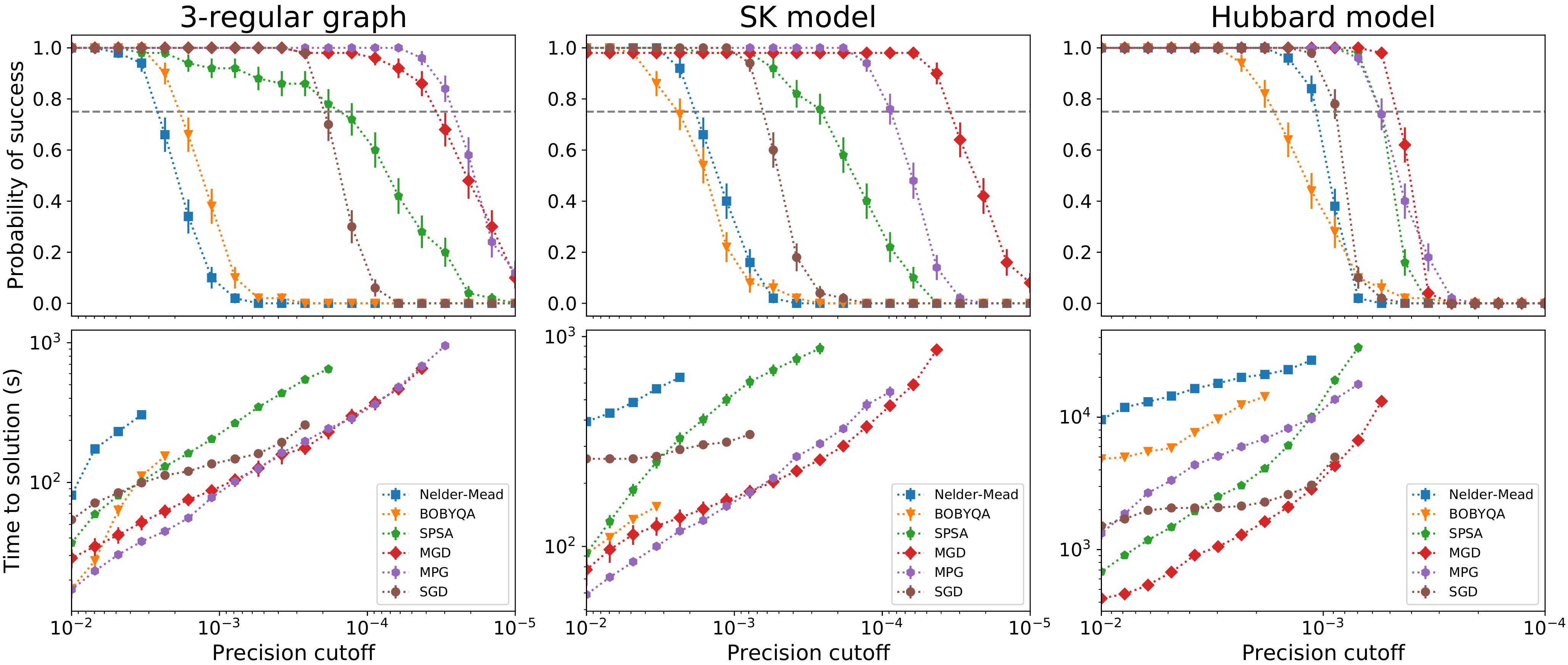}
\caption{Success probability and time to solution for varying levels of required precision at $p=5$.
Top: The probability of converging (out of 50 trials) to the optimal value of the
ansatz at the given precision.
Bottom: The average wall clock time the optimizer took to reach the given precision.
Error bars represent 1 standard deviation. Time to solution is
only reported if the probability of convergence was at least 75\%
(dotted horizontal gray line).
We see that Nelder-Mead and BOBYQA are the least likely
to converge and often the slowest to converge when they do
succeed. Meanwhile, MGD and MPG have the highest probability
of converging as well as usually the fastest convergence times.
}
\label{fig:time_to_solution_n8_p5}
\end{figure*}

\subsection{The case of \texorpdfstring{$p=5$}{p=5} and no gate errors}

At $p=5$ there are a greater number of parameters to optimize.
For the QAOA problems there are now 10 parameters, and for the Hubbard model
there are 15. Here we fixed the wall clock time model to the one that includes
network latency and circuit batching, and plot the performance of the optimizers
as a function of the desired level of precision of convergence to the ansatz optimum.
We present the results
in \fig{time_to_solution_n8_p5}. The optimizers
did not always converge within the time limit we
allowed (1,500 seconds for the QAOA problems and
24 hours for the Hubbard model). 
The top row depicts the probability
of convergence to the desired precision, out of 50 runs. The bottom row depicts the
average wall clock time for convergence, with data plotted only if the probability of
convergence was at least 75\%.

These simulations show that not only were Nelder-Mead and BOBYQA the least likely to converge;
they were also often the slowest to converge when they did succeed. Meanwhile, MGD, MPG, and SPSA converged
even at high levels of precision, with MGD and MPG consistently converging the most quickly in this regime.
This is again a symptom of the fragility of using deterministic optimizers in a stochastic
setting. Outside the regime of precise tuning, methods like Nelder-Mead and BOBYQA become
unstable, whereas even outside the regime of tuning, methods like MGD, MPG, and SPSA are able to succeed.

Note that the plots would look different if we had tuned the hyperparameters
with a different strategy. For example, we tuned the hyperparameters
to minimize the time to convergence to a precision of $10^{-3}$. If we had instead
used a less precise cutoff, such as $10^{-2}$, then we would expect the
optimizers to converge faster to less precise cutoffs, but perhaps more slowly or less robustly to
higher precision cutoffs at smaller ones. At a glance in these figures, one can see
remnants of the hyperparameter selection cutoff.  In Appendix \ref{app:hyperparameters}
we highlight this effect with an example.

\subsection{The impact of rotation errors at \texorpdfstring{$p=5$}{p=5} }

\begin{figure}
\includegraphics[width=1.0\linewidth]{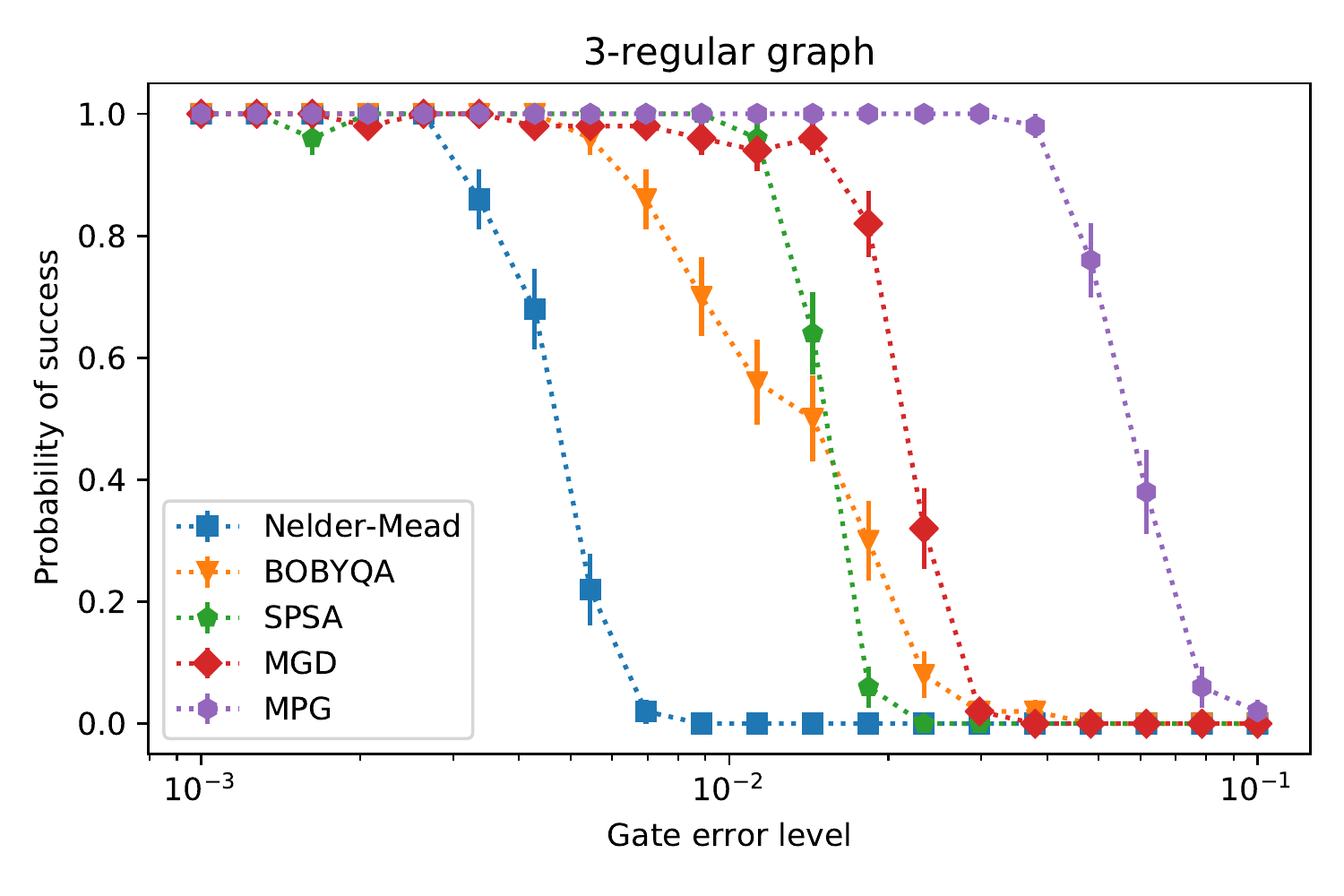}
\caption{Probability of convergence as a function of gate error level
under a model of rotation error for the 3-regular graph model.
Shown is the probability, over 50 trials with different
PRNG seeds, of converging to within a precision of 5e-3,
as a function of gate error level.
Error bars represent one standard deviation.
In this scenario, Nelder-Mead is the least resilient to this noise,
while MPG is the most, and MGD follows.
}
\label{fig:success_probability_vs_noise_level_3reg}
\end{figure}

Finally, to understand the impact of gate error in addition to simple sampling noise, 
at $p=5$ we consider gate rotation errors as well. As described above,
the model of gate rotation error that we used does not simply translate to SGD,
so we do not include results for it.
Again, we fixed the wall clock time model to the one that includes
network latency and circuit batching. 
In running the optimization algorithms, we used the hyperparameters
that were optimized for the case of no gate errors.

\fig{success_probability_vs_noise_level_3reg} shows the probability
of convergence to a precision of $5 \times 10^{-3}$ for the various optimizers
as a function of the gate error level $\varepsilon$, for the
3-regular graph model.
The results show that in this scenario, Nelder-Mead is the least resilient
to this type of noise, while MPG performs the best and MGD follows. The reason why MPG is particularly robust to noise is because it learns a stochastic policy. Its probability-based optimization minimizes the objective function in the expectation sense and thus manages to handle various levels of uncertainty. 
SPSA also showed good noise resilience in other scenarios; see
Section \ref{app:additional_data} in the appendix for data for
the other models.

Note that for a given gate error level,
algorithmic improvements can increase the success probability with respect to the ideal solution
only up to a certain point.  That is, beyond a certain level of noise, the device cannot produce
a more precise solution, and hence this is not a failing of the optimizer but rather represents
a device limitation.  We do not differentiate between these circumstances in the presented data, but
merely note that it is a consideration when defining probability of success.

\section{Conclusion}
\label{sec:conclusion}

Variational quantum algorithms are a promising candidate for execution on near-term
quantum computers, and a number of experimental demonstrations of these algorithms
have already been performed. These algorithms rely on a classical optimization subroutine,
and hence the efficiency of these algorithms can be limited by the performance of these optimizers.
Here, we saw that to accurately assess the performance of these optimizers, it is crucial
to develop a good cost model, and tune available hyperparameters to operational specifications.

Given the unique considerations of quantum systems, we developed two new surrogate model-based optimizers, MGD and MPG, 
to fill some of the gaps of previous methods.  We numerically compared their performance with other
popular alternatives, and found it advantageous in several realistic settings.
We also probed how the cost model and presence of errors can significantly impact the choice
of optimizer in a practical setting. 

Now that quantum computers are
coming online, accessing superconducting qubits through a cloud interface
is an important scenario to consider. The latency of communicating over the
Internet can cause large increases in running times, but this can be mitigated
by circuit batching, though the cost savings depends on the optimizer.

We also observed that inherently stochastic optimizers, such as MPG, MGD and SPSA,
were more robust to variations in problems or setting once properly tuned.  This
extended to situations where finite gate or circuit noise was present.  In contrast, while
it was sometimes possible to make deterministic optimizers competitive through careful tuning,
these tunings were fragile with respect to small variations in the problem or the introduction of
noise. Overall, MPG and MGD's tolerance of noise, ability to take advantage of circuit batching,
and good overall performance make them good candidates for actual experiments,
but the best optimizer can depend on the processor's wall-clock model,
level of noise, number of parameters, or the specific circuit ansatz.

In this work, we have shown how practical considerations can significantly affect
the calculus of choosing an optimizer for running variational algorithms. Future work
will develop more accurate noise and cost models, and further development
of optimizers can take these unique considerations into account.

\subsection*{Acknowledgments}

We thank Eddie Farhi and Bill Huggins for helpful discussions. This work was partially supported by the Department of Energy under Grant No. DE-SC0017867 (J.Y., L.L.) and a Google Quantum Research Award (L.L.).

\subsection*{Code Availability}

Implementations of Model Gradient Descent and Model Policy Gradient are available at \url{https://github.com/quantumlib/ReCirq}.

\bibliography{main}

\onecolumngrid
\appendix
\setcounter{table}{0}
\renewcommand{\thetable}{S\arabic{table}}%
\setcounter{figure}{0}
\renewcommand{\thefigure}{S\arabic{figure}}%

\newpage

\section{Pseudocode for MGD and MPG}
\label{app:pseudocode}

In this section, we give pseudocode for the algorithms Model Gradient Descent (MGD) and Model Policy Gradient (MPG).
The pseudocode for MGD is given in Algorithm \ref{alg:mgd},
and the pseudocode for MPG is given in Algorithm \ref{alg:pg}

\begin{algorithm}[H]
\caption{Model Gradient Descent}\label{alg:mgd}
\begin{algorithmic}[1]
  \Require Initial point $x_0$, learning rate $\gamma$, sample radius $\delta$, sample number $k$,
    rate decay exponent $\alpha$, stability constant $A$,
    sample radius decay exponent $\xi$, tolerance $\varepsilon$, maximum evaluations $n$
 \State Initialize a list $L$
 \State Let $x \leftarrow x_0$
 \State Let $m \leftarrow 0$
 \While{(\#function evaluations so far) + $k$ does not exceed $n$}
  \State Add the tuple $(x, f(x))$ to the list $L$
  \State Let $\delta' \leftarrow \delta / (m + 1)^\xi$
  \State Sample $k$ points uniformly at random from the $\delta'$-neighborhood of $x$;
  Call the resulting set $S$
  \For{each $x'$ in $S$}
    \State Add $(x', f(x'))$ to $L$
  \EndFor
  \State Initialize a list $L'$
  \For{each tuple $(x', y')$ in $L$}
    \If{$\lvert x' - x \rvert < \delta'$}
      \State Add $(x', y')$ to $L'$
    \EndIf
  \EndFor
  \State Fit a quadratic model to the points in $L'$ using least squares
  linear regression with polynomial features
  \State Let $g$ be the gradient of the quadratic model evaluated at $x$
  \State Let $\gamma' = \gamma / (m + 1 + A)^\alpha$
  \If{$\gamma' \cdot \lvert g \rvert < \varepsilon$}
     \State \Return $x$
  \EndIf
  \State Let $x \leftarrow x - \gamma' \cdot g$
  \State Let $m \leftarrow m + 1$
 \EndWhile
 \State \Return $x$
\end{algorithmic}
\end{algorithm}

\begin{algorithm}[H]
\caption{Model Policy Gradient}\label{alg:pg}
\begin{algorithmic}[1]
  \Require learning rate $\gamma$, sample radius ratio $\delta_r$, sample number $k$, model sample number $M$, learning rate decay exponent $\alpha$, mean initialization $\boldsymbol{\mu}_0$, standard deviation initialization $ \sigma_0$, decay steps $t_{\text{decay}}$, warm up steps $t_{\text{warm}}$, maximum evaluations $n$
 \State Initialize a list $L$
 \State Initialize the policy: $\boldsymbol{\mu} \leftarrow \boldsymbol{\mu}_0, \bm{\sigma} \leftarrow [\sigma_0, \cdots, \sigma_0]^T.$
 \State Let $m \leftarrow 0$
 \While{(\#function evaluations so far) + $k$ does not exceed $n$}
  \State Greedy estimation by the current policy: $x \leftarrow \arg \max_{\tilde{x}} {\pi_{\boldphi}} (\tilde{x}).$
  \State Add the tuple $(x, f(x))$ to the list $L$
  \State Sample $k$ points according to the current policy $\pi_{\boldphi}$;
  Call the resulting set $S$
  \For{each $x'$ in $S$}
    \State Add $(x', f(x'))$ to $L$
  \EndFor
  \State Estimate the maximal radius within the set $S$, i.e. $r_{\max{}} 
  \leftarrow \max_{x' \in S} \lvert x' - x \rvert.$
  \If{ $m < t_{\text{warm}}$ }
    \Comment{Compute the policy gradient directly}
    \State Compute the baseline $\bar f \leftarrow \frac{1}{k}\sum_{  x'  \in S } f(x').$
    \State Compute the policy gradient using the sampled data points
    \[ \nabla_{\boldphi} J({\boldphi}) \leftarrow \frac{1}{k}\sum_{  x'  \in S } \nabla_{\boldphi}
  \log \pi_{\boldphi}( x') \cdot (f(x') - \bar f). \]
   \algstore{mpg}
\end{algorithmic}
\end{algorithm}

\clearpage

\begin{algorithm}[H]
  \caption{Model Policy Gradient (continued)}
  \begin{algorithmic}
  \algrestore{mpg}
 \Else
  \Comment{Fit the model to compute the policy gradient}
  \State Initialize a list $L'$
  \For{each tuple $(x', y')$ in $L$}
    \If{$\lvert x' - x \rvert < \delta_r r_{\max}$}
      \State Add $(x', y')$ to $L'$
    \EndIf
  \EndFor
  \State Fit a quadratic model $F(\cdot)$ to the points in $L'$ using least squares
  linear regression with polynomial features
  \State Sample $M$ points according to the current policy $\pi_{\boldphi}$;
  Call the resulting set $S'$
  \For{each $x'$ in $S'$}
    \State Evaluate with the model  $F(x')$.
  \EndFor
\State Compute the baseline $\bar F \leftarrow \frac{1}{M}\sum_{  x'  \in S' } F(x').$
  \State Compute the policy gradient 
      \[ \nabla_{\boldphi} J({\boldphi}) \leftarrow \frac{1}{M}\sum_{  x'  \in S' } \nabla_{\boldphi}
  \log \pi_{\boldphi}( x') \cdot (F(x') - \bar F). \] 
  \EndIf
  \State Decay the learning rate $\gamma' \leftarrow \gamma \cdot \alpha^{m / t_{\text{decay}}}$
  \State Update the weights ${\boldphi} \leftarrow {\boldphi} - \gamma' \cdot \nabla_{\boldphi} J({\boldphi})$
  \State Let $m \leftarrow m + 1$
 \EndWhile
   \State Greedy estimation by the current policy: $x \leftarrow \arg \max_{\tilde x} {\pi_{\boldphi}} (\tilde x).$
 \State \Return $x$
\end{algorithmic}
\end{algorithm}

\section{Hyperparameter selection}
\label{app:hyperparameters}

Each algorithm we studied had hyperparameters and the choice of these hyperparameters had a great
impact on performance. We tuned hyperparameters by performing either a grid search,
or, when this was not feasible, a random search over points on a grid.

\begin{figure}[h]
    \centering
    \includegraphics[width=0.5\linewidth]{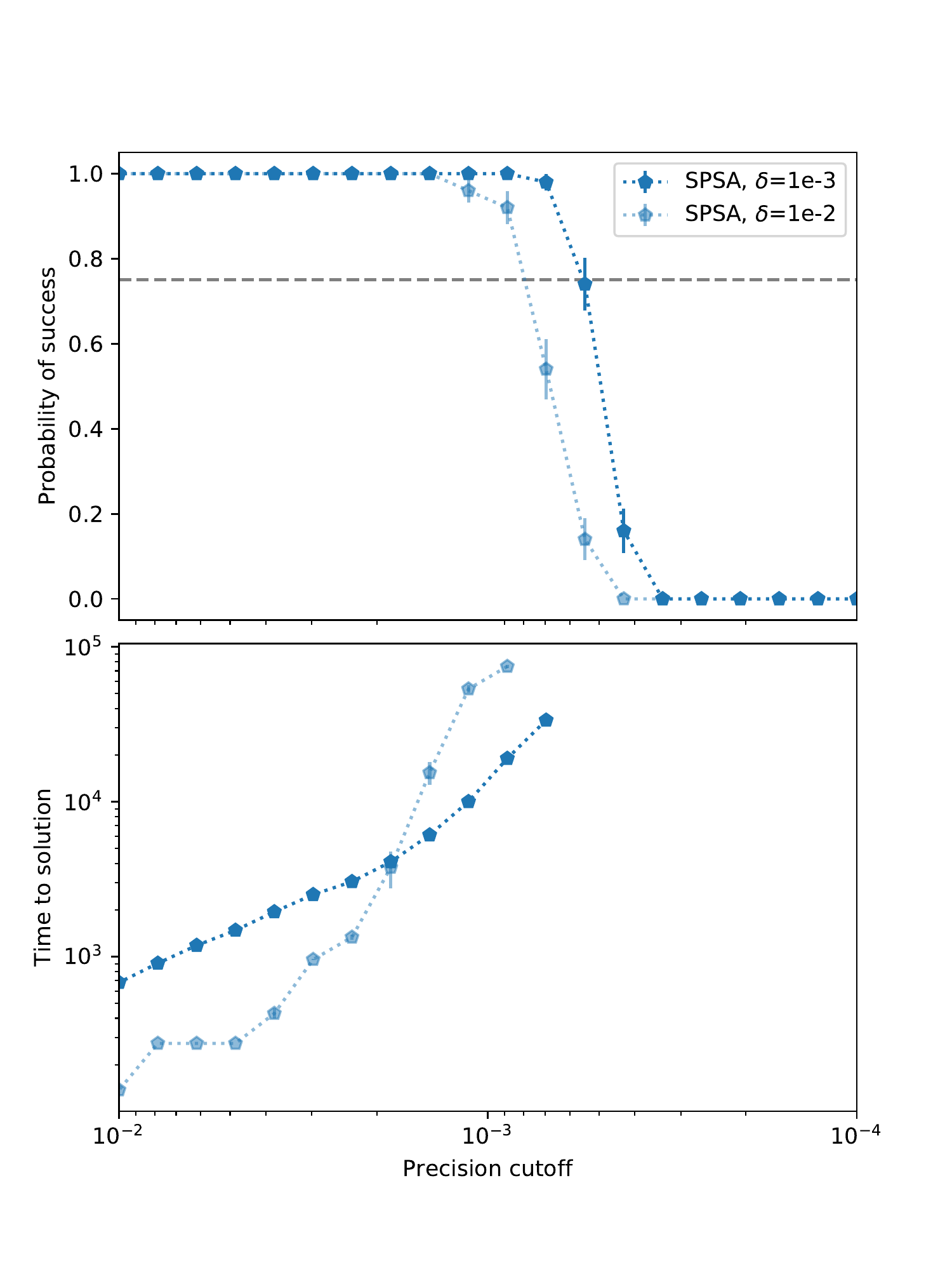}
    \caption{Success probability and time to solution for varying levels of required precision at $p=5$, for SPSA on the Hubbard model.
    Results are shown for two hyperparameter settings, optimized for two
    different precision cutoffs $\delta$:
    $10^{-3}$ (dark colored) and $10^{-2}$ (light colored).
    Top: The probability of converging (out of 50 trials) to the optimal value of the
    ansatz at the given precision.
    Bottom: The average wall clock time the optimizer took to reach the given precision.
    Error bars represent 1 standard deviation. Time to solution is
    only reported if the probability of convergence was at least 75\%
    (dotted horizontal gray line).}
    \label{fig:success_probability_and_time_hyperparameter_comparison}
\end{figure}

For each combination of hyperparameters
considered in the search, we performed an optimization run using the
wall clock time model that includes network latency and circuit batching.
The optimal hyperparameters were those that minimized time to convergence with a precision target of $10^{-3}$. Note that this choice does have an effect on the
performance of the algorithms; choosing a more lenient precision
target would give different results. To demonstrate this effect,
we optimized hyperparameters of SPSA for the Hubbard model
for a precision target of $10^{-2}$ instead of $10^{-3}$.
The results are shown in Figure 
\ref{fig:success_probability_and_time_hyperparameter_comparison}.
As expected, the algorithm optimized for $10^{-2}$ performs better at
larger precision cutoffs and worse at smaller ones.

Below, we describe the hyperparameters of these algorithms and the values that we searched through.
For each algorithm, we considered the number of measurement shots per
energy evaluation to be a hyperparameter, and for each algorithm we considered different sets of
possible values between the QAOA and Hubbard model problems. In the tables below, there is
one line for the values considered for the QAOA problems, and one line for the values
considered for the Hubbard model.
We include tables of the hyperparameters chosen by our grid search.

\subsection{Nelder-Mead}

The Nelder-Mead simplex method has a single additional hyperparameter which we call $\delta$.
This hyperparameter affects the size of the initial simplex. Given an initial guess $\boldtheta_0$,
the algorithm constructs its initial simplex $(\boldtheta_0, \boldtheta_1, \ldots, \boldtheta_m)$,
where $m$ is the dimension
of $\boldtheta_0$, by defining $\boldtheta_i$ to be equal to $\boldtheta_0$ but with its $i$-th coordinate
multiplied by $1 + \delta$. In Table \ref{tab:nm_hyperparameters}
we show the hyperparameter values that we searched through. In Table \ref{tab:nm_hyperparameters_optimized}
we show the hyperparameters that were chosen by the search.

\begin{table}[h]
\begin{tabular}{|c|c|}
    \hline
    Hyperparameter & Possible values \\ \hline \hline
    number of shots (QAOA) & 5,000, 25,000, 125,000, 625,000 \\ \hline
    number of shots (Hubbard) & 10,000, 100,000, 1,000,000, 10,000,000 \\ \hline
    $\delta$ (determines initial simplex size) & 0.001, 0.002, 0.004, 0.008, 0.016, 0.032, 0.064, 0.128, 0.256, 0.512 \\
    \hline
\end{tabular}
\caption{Hyperparameter selection for Nelder-Mead}
\label{tab:nm_hyperparameters}
\end{table}

\begin{table}[h]
\begin{tabular}{|c|c|c|c|c|c|}
    \hline
    Hyperparameter & 3-reg (p=1) & 3-reg (p=5) & SK (p=1) & SK (p=5) & Hubbard (p=5)\\ \hline \hline
    number of shots & 25,000 & 25,000 & 25,000 & 125,000 & 10,000,000 \\ \hline
    $\delta$ & 0.128 & 0.064 & 0.064 & 0.256 & 0.256\\
    \hline
\end{tabular}
\caption{Optimized hyperparameters for Nelder-Mead}
\label{tab:nm_hyperparameters_optimized}
\end{table}

\subsection{Bounded Optimization By Quadratic Approximation}

The BOBYQA algorithm maintains a set of points $(\boldtheta_1, \ldots, \boldtheta_k)$ through which it fits
an interpolating quadratic model. In each iteration, it uses the model to predict a good
point to go next, and incorporates that point into the model by replacing another point.
The model is only assumed to be accurate within a ``trust region radius'' $\rho$ of
the most recently added point.

The value of $k$ is a hyperparameter that can take values from $\{m+1, \ldots, (m+1) (m+2) / 2\}$.
In an $m$-dimensional optimization problem, it takes $(m+1) (m+2) / 2$ points to fully
determine a quadratic function. Thus, if $k$ is smaller than this value, there is some freedom
in choosing the particular quadratic function. BOBYQA takes up this freedom by
minimizing the Frobenius norm of the difference between the Hessians of successive quadratic models.
Instead of using $k$ directly as a hyperparameter, we defined a transformed hyperparameter $\alpha$
taking values from $[0, 1]$ and derived $k$ from it using the formula
$k = \lfloor (m+1) + \alpha [(m+1) (m+2) / 2 - (m+1)] \rfloor$.

BOBYQA also has a hyperparameter we call $\rho_0$ which is the trust region radius at the beginning
of the algorithm. In Table \ref{tab:bobyqa_hyperparameters} we show the hyperparameter
values that we searched through. In Table \ref{tab:bobyqa_hyperparameters_optimized}
we show the hyperparameters that were chosen by the search.

\begin{table}[h]
\begin{tabular}{|c|c|}
    \hline
    Hyperparameter & Possible values \\ \hline \hline
    number of shots (QAOA) & 5,000, 25,000, 125,000, 625,000 \\ \hline
    number of shots (Hubbard) & 10,000, 100,000, 1,000,000, 10,000,000 \\ \hline
    $\alpha$ (determines number of points to interpolate) & 0.0, 0.2, 0.4, 0.6, 1.0 \\ \hline
    $\rho_0$ (initial trust region radius) & 0.01, 0.02, 0.04, 0.08, 0.16 \\
    \hline
\end{tabular}
\caption{Hyperparameter selection for BOBYQA}
\label{tab:bobyqa_hyperparameters}
\end{table}

\begin{table}[h]
\begin{tabular}{|c|c|c|c|c|c|}
    \hline
    Hyperparameter & 3-reg (p=1) & 3-reg (p=5) & SK (p=1) & SK (p=5) & Hubbard (p=5) \\ \hline \hline
    number of shots & 25,000 & 25,000 & 125,000 & 25,000 & 10,000,000 \\ \hline
    $\alpha$ & 0.6 & 0.2 & 1.0 & 0.2 & 0.2 \\ \hline
    $\rho_0$ & 0.04 & 0.16 & 0.08 & 0.16 & 0.04 \\
    \hline
\end{tabular}
\caption{Optimized hyperparameters for BOBYQA}
\label{tab:bobyqa_hyperparameters_optimized}
\end{table}

\subsection{Stochastic gradient descent}

SGD has two additional parameters, the learning rate $\gamma$
and the decay rate $\beta$. These determine the update rule that uses the current gradient $\mathbf{g}_j$
to update the current point $\boldtheta_j$ to the next point $\boldtheta_{j+1}$ as follows:
$$
\boldtheta_{j+1} = \boldtheta_j - \gamma e^{-\beta j} \mathbf{g}_j.
$$
In Table \ref{tab:sgd_hyperparameters} we show the hyperparameter
values that we searched through. In Table \ref{tab:sgd_hyperparameters_optimized}
we show the hyperparameters that were chosen by the search.

\begin{table}[h]
\begin{tabular}{|c|c|}
    \hline
    Hyperparameter & Possible values \\ \hline \hline
    number of shots & 1000, 5000, 10000, 20000, 40000 \\ \hline
    number of shots (Hubbard) & 10,000, 100,000, 1,000,000, 10,000,000 \\ \hline
    $\gamma$ (learning rate) & 0.001, 0.002, 0.004, 0.008, 0.016, 0.032, 0.064, 0.128, 0.256 \\ \hline
    $\beta$ (decay rate) & 0.01, 0.02, 0.04, 0.08, 0.16, 0.32 \\
    \hline
\end{tabular}
\caption{Hyperparameter selection for SGD}
\label{tab:sgd_hyperparameters}
\end{table}

\begin{table}[H]
\centering
\begin{tabular}{|c|c|c|c|c|c|}
    \hline
    Hyperparameter & 3-reg (p=1) & 3-reg (p=5) & SK (p=1) & SK (p=5) & Hubbard (p=5) \\ \hline \hline
    number of shots & 1,000 & 1,000 & 1,000 & 1,000 & 10,000 \\ \hline
    $\gamma$ & 0.016 & 0.008 & 0.008 & 0.004 & 0.004 \\ \hline
    $\beta$ & 0.32 & 0.02 & 0.16 & 0.08 & 0.32 \\
\hline
\end{tabular}
\caption{Optimized hyperparameters for SGD}
\label{tab:sgd_hyperparameters_optimized}
\end{table}

\subsection{Simultaneous Perturbation Stochastic Approximation}

SPSA estimates the gradient $\mathbf{g}_j$ at point $\boldtheta_j$ using the expression
$$
\mathbf{g}_{j, k} = \frac{f(\boldtheta_j + c_j \boldDelta_j) - f(\boldtheta - c_j \boldDelta_j)}{2c_j} \cdot \boldDelta_{j, k}^{-1}
$$
where $\boldDelta_j$ is chosen in each iteration to be a vector whose entries are chosen to be plus
or minus 1 with equal probability and $c_j = c / j^\gamma$ where $c$ and $\gamma$ are hyperparameters
called the perturbation size and perturbation decay exponent, respectively. The new point $\boldtheta_{j+1}$
is calculated according to the update rule
$$
\boldtheta_{j+1} = \boldtheta_j - a_j \mathbf{g}_j
$$
where $a_j = a / (j + A)^\alpha$ where $a$, $\alpha$, and $A$ are hyperparameters called the rate,
rate decay exponent, and stability constant, respectively.

In Table \ref{tab:spsa_hyperparameters} we show the hyperparameter
values that we searched through.
Instead of trying every possible combination, we randomly picked 1000
combinations.
In Table \ref{tab:spsa_hyperparameters_optimized}
we show the hyperparameters that were chosen by the search.

\begin{table}[h]
\begin{tabular}{|c|c|}
    \hline
    Hyperparameter & Possible values \\ \hline \hline
    number of shots (QAOA) & 5,000, 25,000, 125,000, 625,000 \\ \hline
    number of shots (Hubbard) & 10,000, 100,000, 1,000,000, 10,000,000 \\ \hline
    $a$ (rate) & 0.005, 0.01, 0.02, 0.04, 0.08 \\ \hline
    $c$ (perturbation size) & 0.01, 0.02, 0.04, 0.08, 0.16 \\ \hline
    $\alpha$ (rate decay exponent) & 0.1, 0.2, 0.4, 0.8 \\ \hline
    $A$ (stability constant) & 0, 50, 100, 200, 400 \\ \hline
    $\gamma$ (perturbation decay exponent) & 0.01, 0.02, 0.04, 0.08, 0.16 \\
    \hline
\end{tabular}
\caption{Hyperparameter selection for SPSA}
\label{tab:spsa_hyperparameters}
\end{table}

\begin{table}[h]
\begin{tabular}{|c|c|c|c|c|c|}
    \hline
    Hyperparameter & 3-reg (p=1) & 3-reg (p=5) & SK (p=1) & SK (p=5) & Hubbard (p=5) \\ \hline \hline
    number of shots & 1,000 & 25,000 & 25,000 & 25,000 & 1,000,000 \\ \hline
    $a$ & 0.08 & 0.04 & 0.005 & 0.01 & 0.01 \\ \hline
    $c$ & 0.16 & 0.01 & 0.02 & 0.02 & 0.02 \\ \hline
    $\alpha$ & 0.4 & 0.8 & 0.2 & 0.8 & 0.8 \\ \hline
    $A$ & 200 & 50 & 50 & 100 & 100 \\ \hline
    $\gamma$ & 0.04 & 0.01 & 0.04 & 0.02 & 0.16 \\
    \hline
\end{tabular}
\caption{Optimized hyperparameters for SPSA}
\label{tab:spsa_hyperparameters_optimized}
\end{table}

\subsection{Model gradient descent}

The Model Gradient Descent algorithm and its hyperparameters are described in Algorithm
\ref{alg:mgd}. In our study we re-parameterized the hyperparameter $k$, the
sample number, in a similar way to how we re-parameterized the number of interpolation
points in BOBYQA. Instead of using $k$ directly as a hyperparameter, we defined
a transformed hyperparameter $\eta$ being a positive real number and
derived $k$ from it using the formula $k = \eta \cdot (m+1)(m+2)/2$,
where $m$ is the dimension of the optimization problem.

In Table \ref{tab:mgd_hyperparameters} we show the hyperparameter
values that we searched through. Instead of trying every possible combination, we randomly picked 1000
combinations. In Table \ref{tab:mgd_hyperparameters_optimized}
we show the hyperparameters that were chosen by the search.

\begin{table}[h]
\begin{tabular}{|c|c|}
    \hline
    Hyperparameter & Possible values \\ \hline \hline
    number of shots (QAOA) & 5,000, 20,000, 80,000 \\ \hline
    number of shots (Hubbard) & 10,000, 100,000, 1,000,000, 10,000,000 \\ \hline
$\gamma$ (rate) & 0.01, 0.02, 0.04, 0.08, 0.16 \\ \hline
    $\delta$ (sample radius) & 0.01, 0.02, 0.04, 0.08, 0.16 \\ \hline
    $\eta$ (determines sample number) & 0.3, 0.6, 0.9, 1.2 \\ \hline
    $\alpha$ (rate decay exponent) & 0.1, 0.2, 0.4, 0.8 \\ \hline
    $A$ (stability constant) & 0, 50, 100, 200, 400 \\ \hline
    $\xi$ (sample radius decay exponent) & 0.01, 0.02, 0.04, 0.08, 0.16 \\
    \hline
\end{tabular}
\caption{Hyperparameter selection for MGD}
\label{tab:mgd_hyperparameters}
\end{table}

\begin{table}[h]
\begin{tabular}{|c|c|c|c|c|c|}
    \hline
    Hyperparameter & 3-reg (p=1) & 3-reg (p=5) & SK (p=1) & SK (p=5) & Hubbard (p=5) \\ \hline \hline
    number of shots & 1,000 & 5,000 & 1,000 & 5,000 & 100,000 \\ \hline
    $\gamma$ & 0.08 & 0.01 & 0.16 & 0.16 & 0.01 \\ \hline
    $\delta$ & 0.08 & 0.08 & 0.04 & 0.04 & 0.08 \\ \hline
    $\eta$ & 0.9 & 0.3 & 1.2 & 0.3 & 0.6 \\ \hline
    $\alpha$ & 0.4 & 0.4 & 0.8 & 0.8 & 0.4 \\ \hline
    $A$ & 100 & 0 & 100 & 400 & 100 \\ \hline
    $\xi$ & 0.08 & 0.08 & 0.02 & 0.01 & 0.04 \\
    \hline
\end{tabular}
\caption{Optimized hyperparameters for MGD}
\label{tab:mgd_hyperparameters_optimized}
\end{table}

\subsection{Model Policy Gradient}
The Model Policy Gradient (MPG) algorithm (Algorithm
\ref{alg:pg}) parameterizes a Gaussian sampling policy and optimizes in its parameter space. The learnable parameters introduced here are the mean and standard deviation of the policy, i.e. $\boldsymbol{\varphi} = \{\boldsymbol{\mu}, \boldsymbol{\sigma}\}$, where $\boldsymbol{\mu}$ and $\boldsymbol{\sigma}$ have the same dimension as the point $\boldtheta$. Every iteration it samples a batch of data points to estimate the direction (Eqn.~\ref{eqn:pg}) which maximizes the expected total reward (in our case, the reward is the negative ground state energy). One drawback of the vanilla policy gradient (VPG) algorithm~\cite{yao_policy_2020} is that it requires a large batch size to control the variance of estimation. In order to enhance the sample efficiency, we integrate the idea of surrogate model-based optimization with the VPG algorithm. A quadratic model is trained by reusing the history data within some trust region of the current estimation $\boldtheta$. Once we have the model, we can query it to output estimations for any data point within the region. Note that the estimations of these data points have little cost compared with the samples in the beginning. Finally, the policy gradient is applied to improve the policy at the end of each iteration

\begin{equation}
\nabla_{\boldsymbol{\varphi}}J(\boldsymbol{\varphi}) = 
\mathop{ \mathbb E}_{ \boldtheta \sim \mathcal N(\boldsymbol \mu, \boldsymbol \sigma) }
\left[\nabla_{\boldsymbol{\varphi}}\log \pi_{\boldsymbol{\varphi}}( \boldtheta )
\cdot
\big( -f(\boldtheta) \big)
\right].
\label{eqn:pg}
\end{equation}

The hyperparameters of the MPG algorithm are described as follows. The optimizer is chosen to be Adam~\cite{kingma2014adam}, with $\beta_1$, $\beta_2$ being 0.9 and 0.999. The learning rate hyperparameter is $\gamma$ with an exponential decay schedule of rate $\alpha$ for every step $t_{\text{decay}}$. The hyperparameter $\sigma_0$ specifies the initialization for the standard deviation of the  Gaussian policy. The hyperparameter $k$ is the sample batch size at each iteration. The sample radius ratio $\delta_r$ with respect to the maximal radius of the samples determines the trust region in which to fit the model. The hyperparameter $t_{\text{warm}}$ is introduced because in the beginning, the number of data points collected is not adequate enough to fit a good model. Thus we adopt the vanilla policy gradient for the first several iterations before we accumulate enough data points. The hyperparameter $M$ is the model sample number to estimate the policy gradient. It needs to be big enough so that the variance of the estimation is low. In our experiments, we used a constant $M=65536$. Since here we use the quadratic model and Gaussian policy, one can also compute the policy gradient analytically, but implementing it this way would allow plugging in different models. 

In Table \ref{tab:pg_hyperparameters} we show the values of the  hyperparameters that we searched through. Instead of exhausting all possible combinations, we randomly picked 1000
combinations. In Table \ref{tab:pg_hyperparameters_optimized}
we show the hyperparameters that were chosen by the search.

\begin{table}[h]
\begin{tabular}{|c|c|}
    \hline
    Hyperparameter & Possible values \\ \hline \hline
    number of shots (QAOA) & 1,000, 5,000, 20,000 \\ \hline
    number of shots (Hubbard) & 10,000, 100,000, 1,000,000, 10,000,000 \\ \hline
    $\gamma$ (learning rate) & 0.001, 0.005, 0.008, 0.01, 0.02 \\ \hline
    $\alpha$ (learning rate decay exponent) & 0.99, 0.96, 0.93, 0.90 \\ \hline
    $\sigma_0$ (standard deviation initialization) & $\exp(-4.0)$,  $\exp(-5.0)$, $\exp(-6.0)$ \\ \hline
     $k$ (sample number) & 10, 20, 40 \\ \hline
    $\delta_r$ (sample radius ratio) & 1.0, 2.0, 3.0 \\ \hline
    $t_{\text{warm}}$ (warm up steps) & 0, 5, 10 \\
    \hline
\end{tabular}
\caption{Hyperparameter selection for MPG}
\label{tab:pg_hyperparameters}
\end{table}

\begin{table}[h]
\begin{tabular}{|c|c|c|c|c|c|}
    \hline
    Hyperparameter & 3-reg (p=1) & 3-reg (p=5) & SK (p=1) & SK (p=5) & Hubbard (p=5) \\ \hline \hline
    number of shots     & 5,000  & 20,000 & 20,000  & 5,000 & 1,000,000  \\ \hline
    $\gamma$            & 0.02  & 0.005 & 0.02  & 0.005 & 0.01  \\ \hline
    $\alpha$            & 0.99  & 0.96 & 0.99  & 0.93 & 0.90 \\ \hline
    $\sigma_0$          & $\exp(-4.0)$  & $\exp(-4.0)$ & $\exp(-4.0)$  &   $\exp(-4.0)$  & $\exp(-5.0)$ \\ \hline
    $k$                 & 10  & 10 & 10  & 20 & 20 \\ \hline
    $\delta_r$          & 3  & 3 & 2   & 2 & 3 \\ \hline
    $t_{\text{warm}}$   & 10  & 5 & 5  & 0 & 10  \\
    \hline
\end{tabular}
\caption{Optimized hyperparameters for MPG}
\label{tab:pg_hyperparameters_optimized}
\end{table}

\section{Initial state for the Hubbard model.}
\label{app:initial_state}

\begin{figure}[h]
  \centering
  \begin{tikzpicture}
    [vert/.style={circle,draw,minimum size=1mm}]
    \node (0) at (0,0) [vert] {0};
    \node (1) at (2,0) [vert] {1};
    \node (2) at (0,-2) [vert] {2};
    \node (3) at (2,-2) [vert] {3};
    \draw [-] (0) -- (1);
    \draw [-] (2) -- (3);
    \draw [-] (0) -- (2);
    \draw [-] (1) -- (3);
  \end{tikzpicture}
  \caption{Labeling of sites for $2 \times 2$ model.}
  \label{fig:plaquette}
\end{figure}
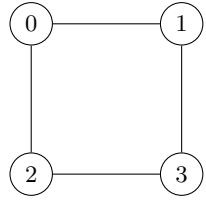

The $2\times 2$ Hubbard model has sites labeled as in \fig{plaquette}.

For a single spin, the single-particle energies of the hopping term are \{-2, 0, 0, 2\}, with corresponding creation operators
\begin{align*}
    b_0^\dagger &= \frac12 (a_0^\dagger + a_1^\dagger + a_2^\dagger + a_3^\dagger) \\
    b_1^\dagger &= \frac{1}{\sqrt{2}}(a_0^\dagger - a_3^\dagger) \\
    b_2^\dagger &= \frac{1}{\sqrt{2}}(a_1^\dagger - a_2^\dagger) \\
    b_3^\dagger &= \frac12 (a_0^\dagger - a_1^\dagger - a_2^\dagger + a_3^\dagger)
\end{align*}

The ground eigenspace is degenerate, and a ground state has the form
\[
    \parens*{\sum_{i, j=1}^2 \alpha_{ij} b_{i, \uparrow} b_{j, \downarrow}} b_{0, \uparrow}^\dagger b_{0, \downarrow}^\dagger \ket{\text{vac}}
\]

Table \ref{tab:hubbard_coeff} lists the choices for the coefficients $\alpha_{ij}$ that give states with the correct total spin (singlet).
\begin{table}[h]
\begin{tabular}{|c||c|c|c|c|}
\hline
Choice & $\alpha_{1,1}$ & $\alpha_{1,2}$ & $\alpha_{2,1}$ & $\alpha_{2,2}$  \\ \hline \hline
1 & 1 & 0 & 0 & 0 \\ \hline
2 & 0 & 0 & 0 & 1 \\ \hline
3 & 0 & $1/\sqrt{2}$ & $1/\sqrt{2}$ & 0 \\ \hline
4 & $1/\sqrt{2}$ & 0 & 0 & $1/\sqrt{2}$ \\ \hline
5 & $1/\sqrt{2}$ & 0 & 0 & $-1/\sqrt{2}$ \\ \hline
\end{tabular}
\caption{Coefficient choices for the $2 \times 2$ Hubbard model
ground state that give the correct total spin.}
\label{tab:hubbard_coeff}
\end{table}

Of these, only choices 3 and 4 led to optimized energies that matched the true ground energy.
We used choice 3 to construct our initial state.

\section{Additional data}
\label{app:additional_data}

\begin{figure}[h]
    \centering
    \includegraphics[width=0.8\linewidth]{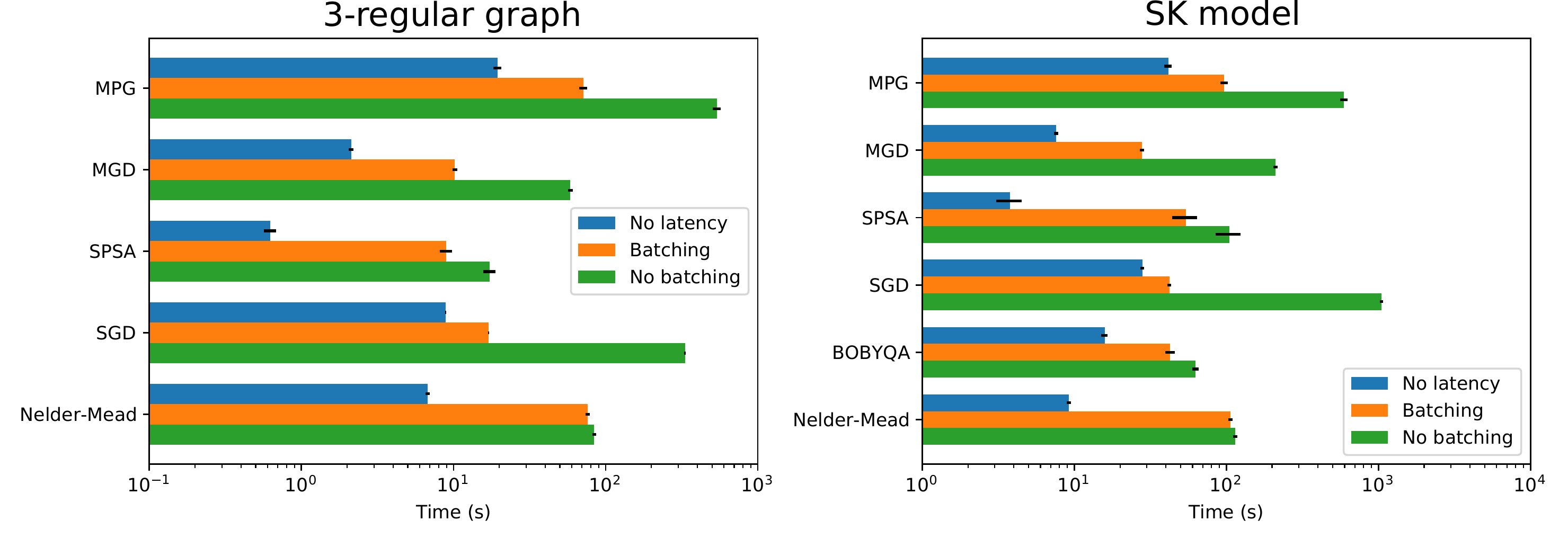}
    \caption{Version of \fig{time_to_solution_sk_n8_p1_NO_NOISE}
    that also includes a plot for the 3-regular graph model.}
    \label{fig:time_to_solution_n8_p1_NO_NOISE} 
\end{figure}

In \fig{time_to_solution_n8_p1_NO_NOISE}
we show a version of \fig{time_to_solution_sk_n8_p1_NO_NOISE}
that also includes a plot for the 3-regular graph model.
For the 3-regular graph model, BOBYQA only converged in 34 out of 50 runs,
so we exclude its data (there other algorithms converged in at least 49
runs).

\begin{figure}[h]
    \centering
    \includegraphics[width=0.8\linewidth]{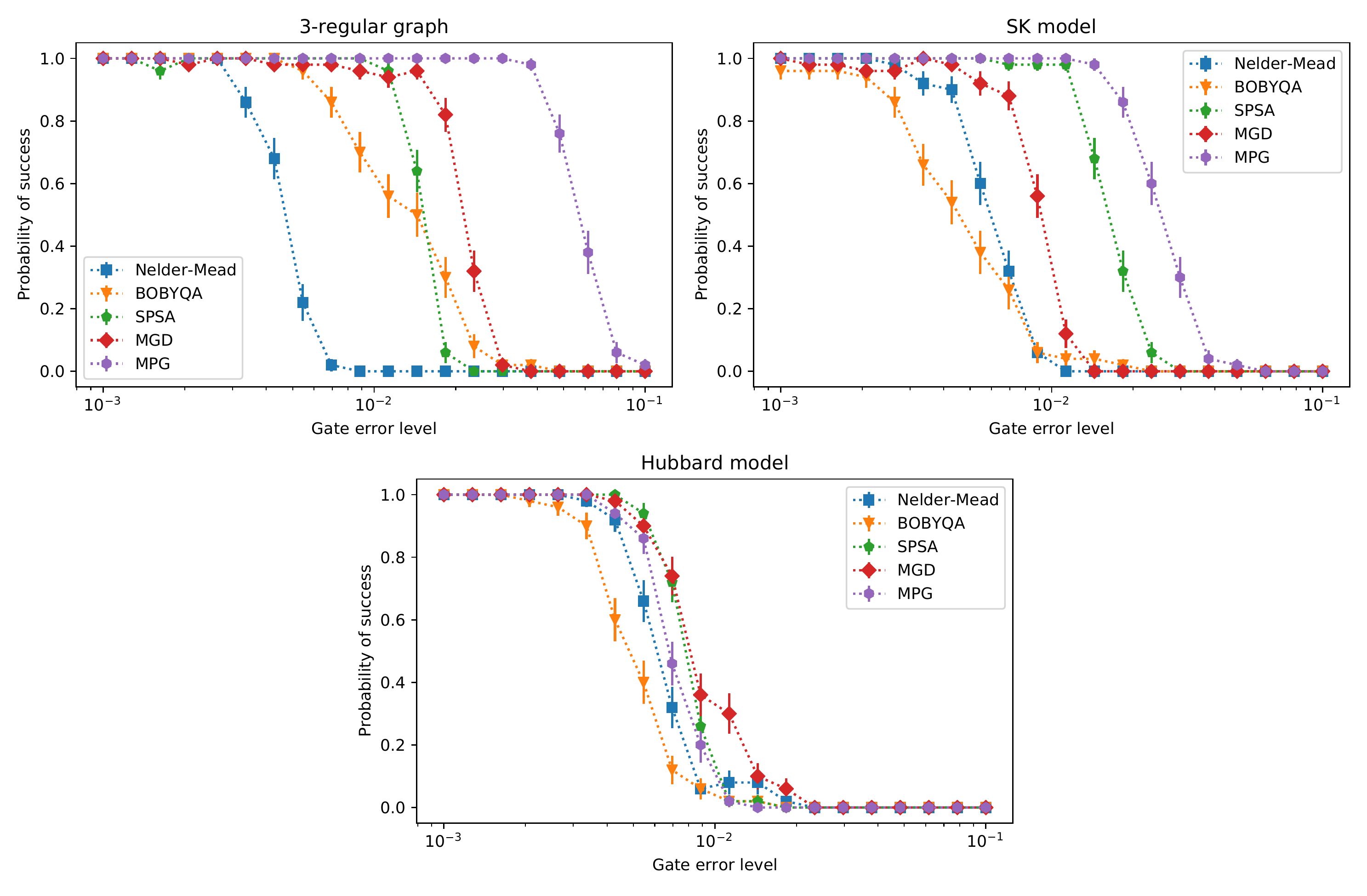}
    \caption{Version of \fig{success_probability_vs_noise_level_3reg}
    that also includes plots for the Sherrington-Kirkpatrick
    and Hubbard models.}
    \label{fig:success_probability_vs_noise_level}
\end{figure}

\begin{figure}[h]
    \centering
    \includegraphics[width=0.8\linewidth]{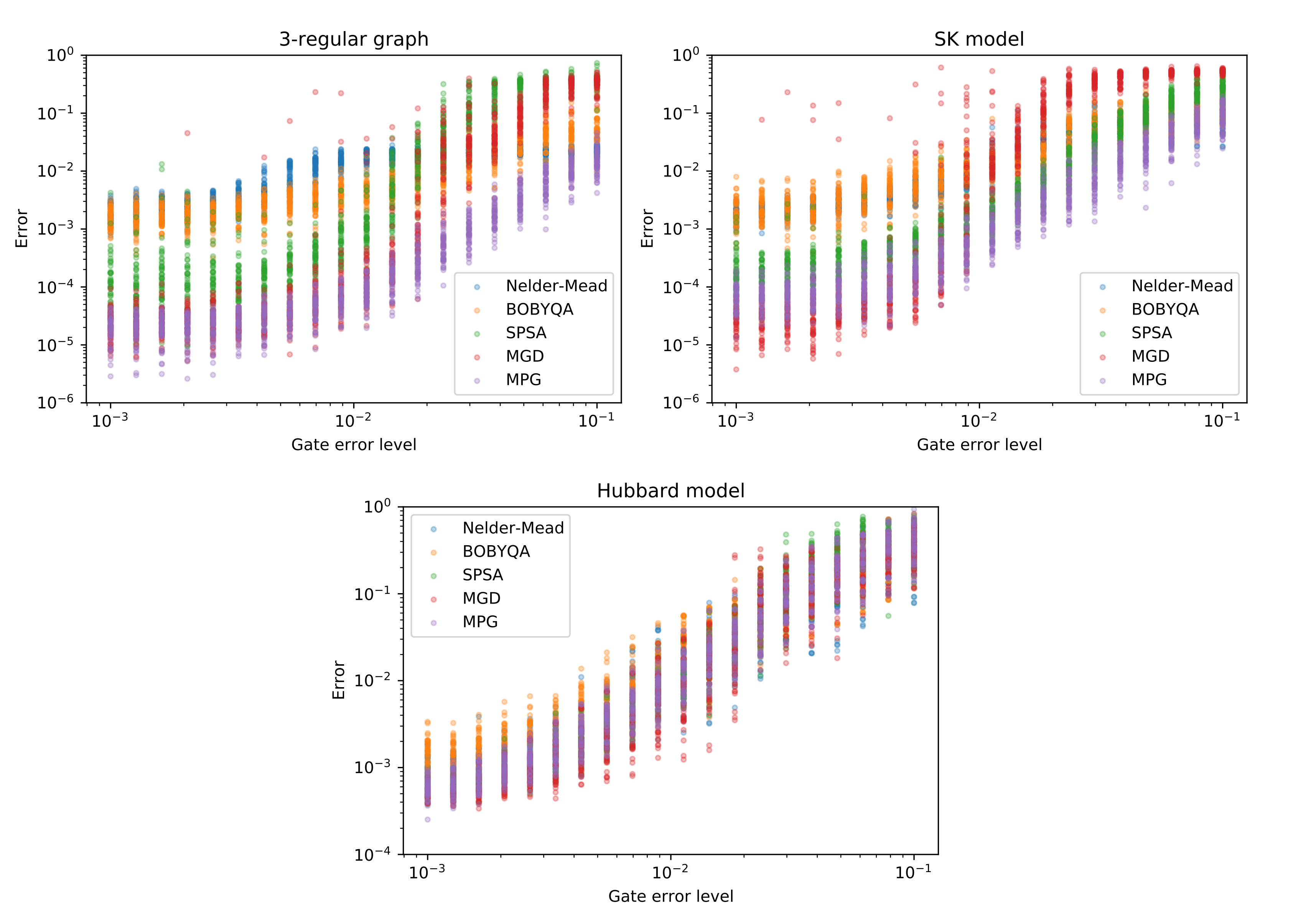}
    \caption{Final energy error as a function of gate error level (amount
    of gate rotation error), for $p=5$. For each gate error level and algorithm,
    the final error for the 50 runs with different PRNG seeds are plotted.}
    \label{fig:error_vs_noise_level}
\end{figure}

\fig{success_probability_vs_noise_level}
shows a version of \fig{success_probability_vs_noise_level_3reg}
that also includes plots for the Sherrington-Kirkpatrick and Hubbard models.
\fig{error_vs_noise_level} plots the final energy error
of the optimizers as a function of the amount of gate rotation
error present, at $p=5$. It shows that for the QAOA problems,
MGD and SPSA clearly outperform the others when the final
energy error is required to be less than about 1e-2.
For the Hubbard model, the optimizers do not differentiate
as clearly.

\section{Optimization trajectories}

\begin{figure}
    \centering
    \includegraphics[width=0.6\linewidth]{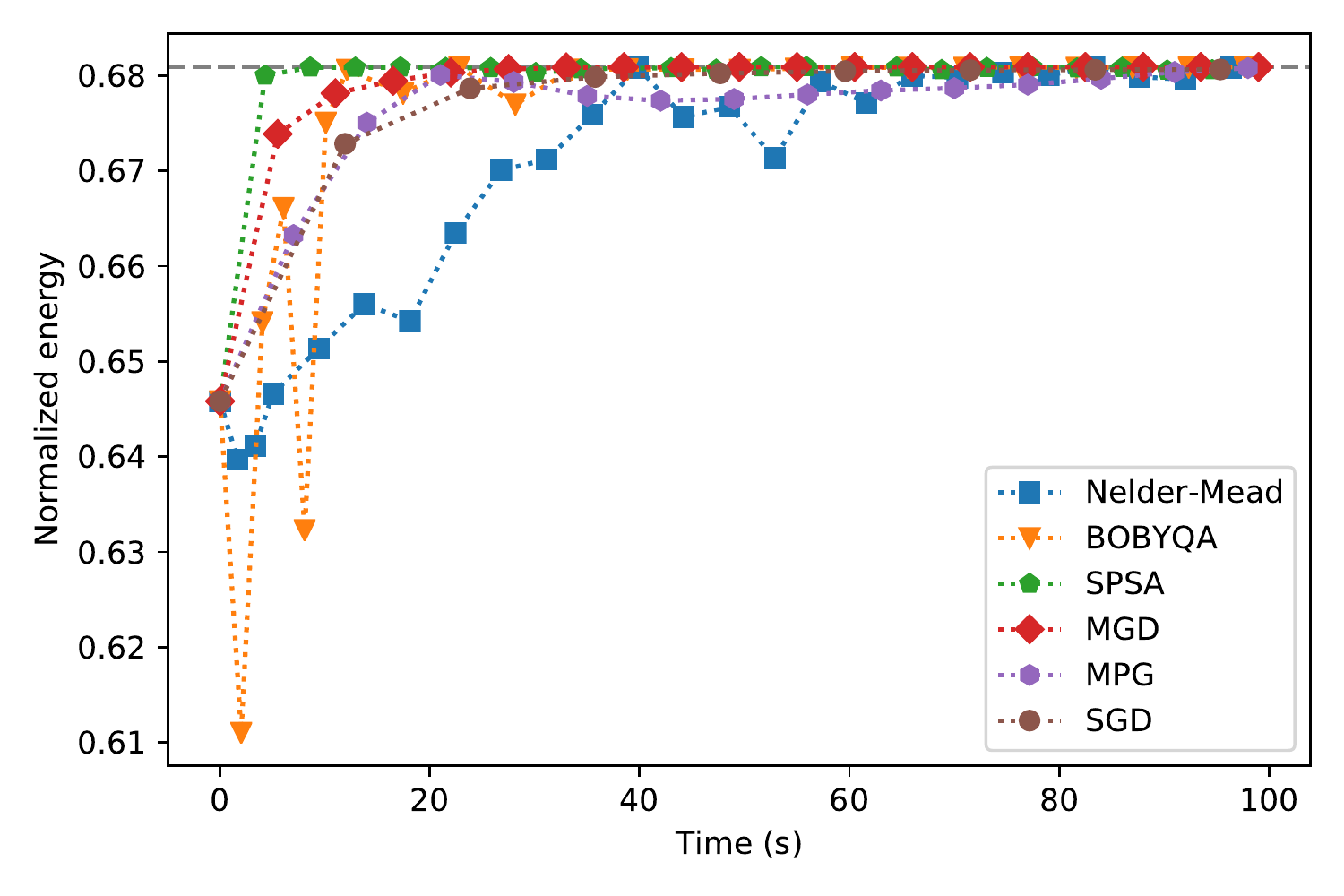}
    \caption{Optimization progress of the optimizers in a simulated experiment on a Sherrington-Kirkpatrick
    model Hamiltonian. Depicted is the normalized energy
    versus wall clock time using the wall clock time model that includes network latency and circuit batching.
    Our use of wall clock time for the x-axis enables a fair comparison to be made between realistic costs.}
    \label{fig:energy_vs_time_example} 
\end{figure}

\fig{energy_vs_time_example} shows the optimization progress of the different optimizers in a simulated
experiment on the Sherrington-Kirkpatrick model with $n=8$ and $p=1$
and no gate errors, with wall clock time
measured using the cost model that includes network latency and circuit batching.
The energy plotted
is the exact expectation value of the quantum state obtained from the parameters
being considered by the optimizer.
The use of wall clock time for the x-axis enables
a fair comparison to be made
between realistic costs. This plot illustrates some differences
between how the optimizers work. MGD, SGD, and SPSA generally
show monotonic progress towards the solution, as does BOBYQA
once it has queried enough points to construct its surrogate model.
On the other hand, the MPG and Nelder-Mead do not
show monotonic progress.

\end{document}